\documentclass[aps,pra,nofootinbib,twocolumn]{revtex4-2}
\usepackage{amsmath, amsthm, amscd, amssymb}
\usepackage{bm}
\usepackage{bbm}
\usepackage{epsfig}

\begin{document}

\title{Minkowski vacuum entanglement and accelerated oscillator chains}
\author{Anatoly A. Svidzinsky$^1$, Marlan O. Scully$^{1,2,3}$, and William
Unruh$^{1,4}$}
\affiliation{$^1$Texas A\&M University, College Station, TX 77843; \\
$^2$Princeton University, Princeton, NJ 08544, USA; \\
$^3$Baylor University, Waco, Texas 76798, USA; \\
$^4$University of British Columbia, Vancouver, Canada V6T 1Z1 }

\date{\today }

\begin{abstract}
Minkowski vacuum is empty from the perspective of Unruh-Minkowski photons,
however, in the Rindler picture, it is filled with entangled pairs of
Rindler photons. A ground-state atom uniformly accelerated through Minkowski
vacuum can become excited by absorbing a Rindler photon (Unruh effect) or,
in the alternative description, by emitting an Unruh-Minkowski photon
(Unruh-Wald effect). We find an exact solution for the quantum evolution of
a long chain of harmonic oscillators accelerated through Minkowski vacuum
and for two chains accelerated in the opposite directions. We show how
entanglement of Rindler photons present in Minkowski vacuum is transferred
to the oscillators moving in causally disconnected regions. We also show
that in the Unruh-Minkowski photon picture the process can be interpreted as
if initial correlations between collective oscillator modes are transferred
to the generated Unruh-Minkowski photons.
\end{abstract}

\maketitle

\section{Introduction}

If we disregard gravity, the geometry of our universe is a four-dimensional
Minkowski spacetime. The Minkowski vacuum is the ground state of a free
field in Minkowski spacetime. Such a state of the field does not excite an
atom moving with a constant velocity through the vacuum for a long time. If
we choose plane-wave modes%
\begin{equation*}
\phi _{\nu }(t,\mathbf{r})=\frac{e^{-i\nu t+i\mathbf{k}\cdot \mathbf{r}}}{%
\sqrt{2(2\pi )^{d}\nu }},\quad \nu =kc>0,
\end{equation*}%
as a basis set to quantize the field, the Minkowski vacuum looks empty,
where $d$ is dimension of space. These modes are positive norm modes (i.e.
modes which go as $e^{-i\nu t}$ with $t$ the Minkowski time and $\nu >0$)
with the norm 
\begin{equation*}
\left\langle \phi _{1},\phi _{2}\right\rangle =\frac{i}{2}\int \left( \phi
_{1}^{\ast }\pi _{2}^{\ast }-\pi _{1}\phi _{2}\right) d^{d}x\mathbf{,}
\end{equation*}%
where $\pi =(1/c)\partial \phi ^{\ast }/\partial t$ is the conjugate
momentum to the field $\phi $. In this paper we will limit ourself to $d=1$,
but the same results apply in the higher dimensions. These modes can also be
written in terms of Unruh-Minkowski modes defined in Eqs. (\ref{UM}) and (%
\ref{UM2}), which is another basis for the usual Minkowski positive norm
modes. Positive norm modes are associated with annihilation operators of the
field.

Atoms interact with the field locally, that is an atom feels the local value
of the field at the atom's location. As a consequence, the local properties
of the photon mode function determine the atom's ability to emit and absorb
a photon. An atom uniformly accelerated in the Minkowski vacuum can become
excited by emitting a photon into the Unruh-Minkowski mode (see Eqs. (\ref%
{UM}) and (\ref{UM2}) below). We call this the Unruh-Wald effect \cite%
{Unru84}. From the perspective of a uniformly accelerated atom, a certain
type of the Unruh-Minkowski modes act as if they have negative frequency and
the atom can become excited by emitting a photon into these locally
\textquotedblleft negative frequency\textquotedblright\ modes \cite{Svid21}.

The positive norm Unruh-Minkowski modes are a complete set of modes which
cover the same set of solutions as do the positive-frequency plane-wave
modes mentioned above. As in the case of plane-wave photons \cite{Scul03},
there are no Unruh-Minkowski photons in the Minkowski vacuum. However, field
quantization in terms of Rindler modes \cite{Rind66} yields that the
Minkowski vacuum if filled with Rindler photons in a two-mode squeezed state
(\ref{w0}) in which the number of photons in different modes is correlated.
A uniformly accelerated atom can become excited by absorbing a Rindler
photon (Unruh effect \cite{Full73,Davi75,Unru76}).

Entanglement (correlations) of the Minkowski vacuum is a subject of a
long-standing interest (it is implicit already in Ref. \cite{Unru84}). Later
it was realized that the vacuum state of quantum field theory can maximally
violate Bell's inequalities \cite{Summ87}. In particular, in 1991, Valentini
showed that a pair of initially uncorrelated bare atoms, separated by a
distance $R$, develop non-local statistical correlations in a time $t<R/c$ 
\cite{Vale91}. Reznik has explored the entanglement of the vacuum of a
relativistic field by letting a pair of causally disconnected probes
interact with the field. He found that, even when the probes are initially
non-entangled, they can wind up in entangled state \cite{Rezn03,Rezn05}.

Recently, there is a growing interest in studing entanglement harvesting -
the process of entangling detectors through their independent interaction
with the field \cite{Salt15,Poza15}. It is more appropriate to call this
process entanglement transfer. This phenomenon, originally investigated in
flat spacetime \cite{Vale91,Rezn03}, has been studied in detail in various
contexts, such as cosmological settings \cite{Stee09,Mart12}, noninertial
motion \cite{Salt15,Liu22,Liu23}, black hole spacetimes \cite%
{Tjoa20,Gall21,Ng22}, and in the presence of gravitational waves \cite%
{Xu20,Gray21}. Effects due to different properties of the detectors, as well
as the dimensions, curvature, and topology of the spacetime, were
investigated \cite{Poza16,Tjo20,Gall21,Sury22,Yan22}. It was found that
entanglement harvesting is quite sensitive to spacetime dimensionality \cite%
{Poza15,Yan22}, topology \cite{Marti16}, curvature \cite%
{Stee09,Namb13,Kuki17,Ng18,Gall21}, and the presence of boundaries \cite%
{Cong19,Cong20,Svid24}. Entanglement of the relativistic vacuum state has a
role in Unruh acceleration radiation \cite{Full73,Davi75,Unru76} and
radiation of evaporating black holes \cite{Hawk74,Hawk75,Svid23}.

In this paper we study dynamics of the Unruh effect by considering a chain
of harmonic oscillators accelerated through the Minkowski vacuum and two
chains accelerated in the opposite directions in the causally disconnected
regions. We find an exact analytical solution for the evolution of the
system's (oscillators + field) state vector in the limit of a long chain. As
we show, in this limit, there are collective excitations of the oscillators
in the chain which are coupled with only one Rindler mode of the field and
system's evolution is factorized.

We find that such extended system undergoes Rabi oscillations and there is
periodic exchange of entanglement between Rindler photons present in the
Minkowski vacuum and the oscillators. At resonance the amplitude of Rabi
oscillations is maximum and there is a complete transfer of entanglement.
Further away from the resonance frequency the oscillation amplitude becomes
smaller and there is partial transfer of entanglement. We also show that in
the Unruh-Minkowski photon picture the process can be interpreted as if
initial correlations between collective modes of the ground-state
oscillators are transferred back and forth to the generated Unruh-Minkowski
photons.

\begin{figure}[h]
\begin{center}
\epsfig{figure=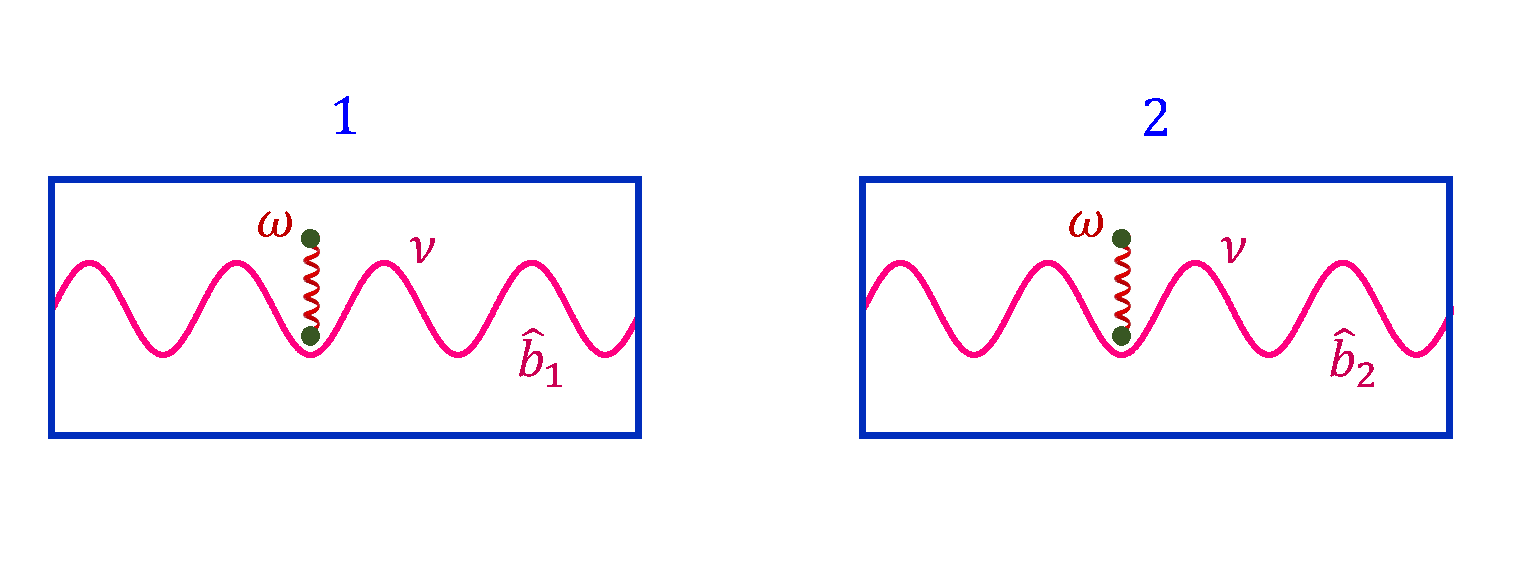, angle=0, width=9cm}
\end{center}
\caption{Harmonic oscillators of frequency $\protect\omega $ are placed
inside identical stationary cavities. If initially cavities are excited such
that number of photons in the cavities are correlated the ground-state
oscillators become excited in a correlated fashion even though cavities are
disconnected.}
\label{Fig0}
\end{figure}

The physics of the effect is somewhat similar to the correlated excitation
of two harmonic oscillators of frequency $\omega $ placed inside entangled
optical cavities shown in Fig. \ref{Fig0}. The cavities and the oscillators
are stationary and the cavities are not connected with each other. Photons
in the cavities have frequencies $\nu _{n}$ and their mode functions 
\begin{equation*}
\phi _{1n}(t,z)=\phi _{1n}(z)e^{-i\nu _{n}t},\quad \phi _{2n}(t,z)=\phi
_{2n}(z)e^{-i\nu _{n}t},
\end{equation*}%
are localized in the cavity 1 and 2 respectively. The corresponding photon
annihilation operators we denote as $\hat{b}_{1n}$ and $\hat{b}_{2n}$, while
annihilation operators for the oscillators are $\hat{\sigma}_{1}$ and $\hat{%
\sigma}_{2}$. We will take the field to be scalar, but modes could equally
be modes of an electromagnetic field.

While the usual vacuum state for the field would be the state $\left\vert
0\right\rangle $ such that 
\begin{equation*}
\hat{b}_{1n}\left\vert 0\right\rangle =\hat{b}_{2n}\left\vert 0\right\rangle
=0
\end{equation*}%
for all $n$, we will instead choose the state based on the squeezed operators%
\begin{equation}
\hat{a}_{1n}=\frac{\hat{b}_{1n}-\gamma \hat{b}_{2n}^{\dag }}{\sqrt{1-\gamma
^{2}}},\quad \hat{a}_{2n}=\frac{\hat{b}_{2n}-\gamma \hat{b}_{1n}^{\dag }}{%
\sqrt{1-\gamma ^{2}}},
\end{equation}%
where $\gamma $ is a real parameter obeying condition $|\gamma |<1$ (in
studies of squeezed states the parameter $\gamma $ is usually written as $%
\gamma =\tanh r$, where $r$ is squeezing parameter which in general case
could depend on the mode). Namely, we assume that state of the field $%
\left\vert 0_{S}\right\rangle $ is vacuum for the squeezed operators $\hat{a}%
_{1n}$ and $\hat{a}_{2n}$. That is%
\begin{equation*}
\hat{a}_{1n}\left\vert 0_{S}\right\rangle =\hat{a}_{2n}\left\vert
0_{S}\right\rangle =0
\end{equation*}%
for all $n$. Mode functions of photons $\hat{a}_{1n}$ and $\hat{a}_{2n}$ are
a superposition of the cavity modes $\phi _{1n}$ and $\phi _{2n}$ with the
opposite sign of frequency 
\begin{equation}
F_{1n}(t,z)=\frac{\phi _{1n}(z)e^{-i\nu _{n}t}-\gamma \phi _{2n}^{\ast
}(z)e^{i\nu _{n}t}}{\sqrt{1-\gamma ^{2}}},  \label{F1}
\end{equation}%
\begin{equation}
F_{2n}(t,z)=\frac{\phi _{2n}(z)e^{-i\nu _{n}t}-\gamma \phi _{1n}^{\ast
}(z)e^{i\nu _{n}t}}{\sqrt{1-\gamma ^{2}}}.  \label{F2}
\end{equation}

We now place a detector (harmonic oscillator of frequency $\omega $) into
the cavity 1 at the location $z_{1}$. The interaction between the oscillator
and the field is governed by the Hamiltonian%
\begin{equation*}
\hat{V}=\hslash g\left( \hat{\sigma}_{1}^{\dag }e^{i\omega t}+\hat{\sigma}%
_{1}e^{-i\omega t}\right) \frac{\partial }{\partial t}\hat{\Phi}(t,z_{1}),
\end{equation*}%
where

\begin{equation*}
\hat{\Phi}(t,z)=\sum_{n}\left( F_{1n}^{\ast }(t,z)\hat{a}_{1n}^{\dag
}+F_{2n}^{\ast }(t,z)\hat{a}_{2n}^{\dag }+h.c.\right)
\end{equation*}%
is the field operator and $g$ is the oscillator-field coupling constant.
Using Eqs. (\ref{F1}) and (\ref{F2}) we obtain%
\begin{equation*}
\hat{V}=\frac{i\hslash g}{\sqrt{1-\gamma ^{2}}}\left( \hat{\sigma}_{1}^{\dag
}e^{i\omega t}+\hat{\sigma}_{1}e^{-i\omega t}\right) \times
\end{equation*}%
\begin{equation*}
\sum_{n}\nu _{n}\left[ \phi _{1n}^{\ast }(z_{1})e^{i\nu _{n}t}\hat{a}%
_{1n}^{\dag }+\gamma \phi _{1n}(z_{1})e^{-i\nu _{n}t}\hat{a}_{2n}^{\dag
}-h.c.\right] .
\end{equation*}%
We assume that initially the oscillator is in its ground state $\left\vert
G\right\rangle $, the field is in the state $\left\vert 0_{S}\right\rangle $
and the oscillator frequency $\omega $ is equal to one of the frequencies of
the cavity modes. To the lowest order%
\begin{equation*}
\hat{V}\left\vert 0_{S}\right\rangle \left\vert G\right\rangle =\frac{%
i\hslash g}{\sqrt{1-\gamma ^{2}}}\hat{\sigma}_{1}^{\dag }\left\vert
G\right\rangle \times
\end{equation*}%
\begin{equation*}
\sum_{n}\nu _{n}\left[ \phi _{1n}^{\ast }(z_{1})e^{i(\omega +\nu _{n})t}\hat{%
a}_{1n}^{\dag }+\gamma \phi _{1n}(z_{1})e^{i(\omega -\nu _{n})t}\hat{a}%
_{2n}^{\dag }\right] \left\vert 0_{S}\right\rangle .
\end{equation*}

Over short times, there will be many frequencies in the field that will
respond to the presence of the detector. Over a time scale much longer than $%
L/c$, where $L$ is the cavity length, the non-resonant terms will average
out to zero, which means that the surviving term is the one associated with
the oscillator being excited to its one quanta state, and the field is
excited into its squeezed state associated with the operator $\hat{a}_{2n}$
for which $\nu _{n}=\omega $.

According to Eq. (\ref{F2}), the amplitude of the mode $F_{2n}$ in the
cavity 2 is $1/|\gamma |>1$ higher than in the cavity 1. Thus because of the
initial squeezed state one has placed the field in, although the detector in
cavity 1 was the one to become excited, it is the field in the cavity 2 that
is excited to a greater extent than the field in the cavity 1. This is all
because of the initial \textquotedblleft vacuum\textquotedblright\
(squeezed) state of the two regions which do not communicate with each
other. For longer time periods, the system will undergo Rabbi oscillations,
as one would expect for a system with just two coupled harmonic oscillators.

One of the surprising features of this model is that, if one were to place a
second detector $\hat{\sigma}_{2}$ into the cavity 2, sensitive to the same
frequency, then the probability that detector 2 would also be excited, if
oscillator 1 was excited would be larger than the product of the
probabilities that oscillator 1 or 2 on their own would be excited. The
field in the cavity 2 is larger than in the cavity 1 because detector in the
cavity 1 (oscillator) is measured to be excited. If the detector $\hat{\sigma%
}_{1}$ is measured to be unexcited then the excitation amplitude of the
detector $\hat{\sigma}_{2}$ is negative. I.e., the number of photons in the
cavity 2 is independent of the presence of the detector in the cavity 1.

During system evolution, correlations between cavity photons $\hat{b}_{1}$
and $\hat{b}_{2}$ in the state $\left\vert 0_{S}\right\rangle $ are
transferred to the oscillators $\hat{\sigma}_{1}$ and $\hat{\sigma}_{2}$. As
a result, oscillators placed in the disconnected cavities become entangled.
If oscillators are in resonance with the cavity mode ($\nu =\omega $) the
system's state vector undergoes Rabi oscillations and entanglement is
transferred back and forth between the field of the cavities and the
oscillators. These Rabi oscillations are a feature of the cavity QED when
the single-mode approximation is accurate.

The oscillator in the cavity 1 can become excited by emitting a photon into
the mode $F_{2}$ which has negative frequency from the oscillator's
perspective because $F_{2n}(t,z_{1})\propto e^{i\nu _{n}t}$. Detection of
the oscillator excitation collapses the state vector of the field into a
state with one photon in the mode $F_{2}$. If $\gamma \ll 1$, this process
mainly affects the field in the other cavity (cavity 2) because, according
to Eq. (\ref{F2}), the mode function $F_{2}$ is predominantly localized in
this cavity. The photon $F_{2}$ has a positive frequency from the
perspective of the oscillator in the cavity 2. Hence, the second oscillator
can become excited by absorbing the photon $F_{2}$. This yields correlated
excitation of oscillators in the cavities 1 and 2.

The systems behave very much as if that second detector had absorbed the
particle created by the first detector (or vice versa) even though there was
no way that the field in the cavity 2 could have been influenced by a change
in the cavity 1. Both cavities are isolated from each other by the assumed
perfectly reflecting walls of the boxes containing the fields in regions 1
and 2.

Moreover, if both detectors were placed into cavity 1, then the probability
that both detectors would end up excited, and the field were left in its
squeezed ground state $\left\vert 0_{S}\right\rangle $ is zero. I.e., the
naive expectation that detector 2 could absorb the particle emitted by the
detector 1 or vice versa, is wrong. However, detector 1's emission could
still influence detector 2 by enhancing the probability that detector 2
emits a particle. Thus the probability that both are excited is enhanced
over the product of the probability that each would do so on its own, a la
Dicke superradiance.

While the above may seem somewhat pedantic and artificial, it is actually
closely related to a standard phenomenon, called the Unruh effect. There, in
flat spacetime, one can introduce two different descriptions of the same
system. By going from the standard flat spacetime Minkowski coordinates to
so called Rindler coordinates, one has the equivalent of the two enclosed
boxes of the above system. 

The toy model of Fig. \ref{Fig0} also demonstrates that Hawking radiation is
not tunneling. Clearly there is no tunneling between isolated cavities of
Fig. \ref{Fig0} but they behave just like being separated by the Rindler
horizon. Tracing over the field in the cavity 1 leaves the field in the
cavity 2 in a thermal state, and vice versa.

In this paper we consider quantum evolution of accelerated harmonic
oscillators coupled to a field in free space and show that it is very
similar to the simple system of two oscillators placed in stationary
cavities we just mentioned. One should note that Rabi oscillations are not
present if an oscillator is interacting with many modes of the field, e.g.,
when a single point-like oscillator is accelerated through the Minkowski
vacuum. We will discuss this case elsewhere. In the present paper, instead
of a single oscillator, we consider a chain of oscillators accelerated in
1+1 dimension. Such a chain acts to form a new \textquotedblleft
cavity\textquotedblright\ \cite{Svid08} yielding Rabi oscillation dynamics.
The accelerated oscillator chain is a better tool to study the Minkowski
vacuum than a single oscillator because field has many modes, and we need a
many-mode device to be able to map modes of the field into the modes of the
device.

\section{Dynamics of the Unruh effect}

\begin{figure}[h]
\begin{center}
\epsfig{figure=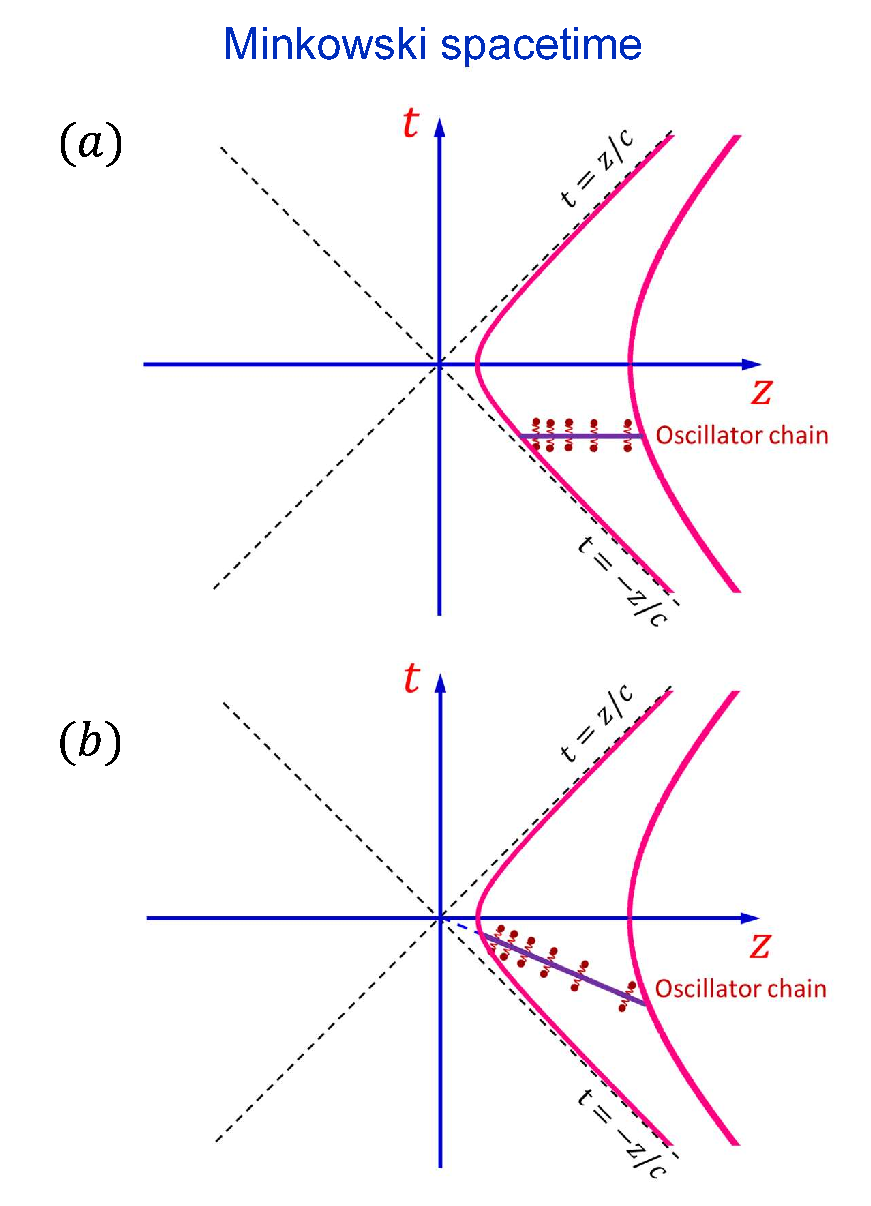, angle=0, width=8cm}
\end{center}
\caption{ Harmonic oscillator chain is accelerated in the right Rindler
wedge through Minkowski vacuum. (a) The chain position is shown at fixed $t$%
. (b) The chain position is shown when the Rindler time $\protect\tau $ is
the same for all oscillators. In this case the chain is a straight line in
the Minkowski spacetime $ct=z\tanh (a\protect\tau /c)$ whose continuation
passes through the origin of the coordinate system. The slope of the line
increases with $\protect\tau $ and the line rotates around the origin during
the chain motion. The linear density of the oscillators in the chain in the
Minkowski spacetime is inversely proportional to the distance to the origin.
In the present discussion we assume a semi-infinite chain which extends from
the horizon $z=c|t|$ to $z=+\infty $. }
\label{Fig1}
\end{figure}

In this Section we consider a chain of identical harmonic oscillators in 1+1
dimension accelerated in the right Rindler wedge through the Minkowski
vacuum $\left\vert 0_{M}\right\rangle $ (see Fig. \ref{Fig1}). We assume
that oscillators in the chain are uniformly accelerated along the
trajectories 
\begin{equation}
t(\tau )=\frac{c}{a}e^{a\bar{z}/c^{2}}\sinh \left( \frac{a\tau }{c}\right)
,\quad z(\tau )=\frac{c^{2}}{a}e^{a\bar{z}/c^{2}}\cosh \left( \frac{a\tau }{c%
}\right) ,  \label{u1}
\end{equation}%
where $\tau $ is the Rindler time and $\bar{z}$ is a parameter describing
position of the oscillator in the chain. Namely, $\bar{z}$ is the oscillator
coordinate in the co-moving frame (Rindler space \cite{Rind66}). We assume
that in the Rindler time the oscillators have the same frequency $\omega $
and are distributed continuously with uniform density. Thus, in Minkowski
space the linear density of the oscillators goes as $|z|/(z^{2}-c^{2}t^{2})$.

In Eq. (\ref{u1}), $a>0$ is a proper acceleration of the oscillator located
at $\bar{z}=0$. An oscillator located at $\bar{z}\neq 0$ is uniformly
accelerated in Minkowski space $(t,z)$ with the proper acceleration $ae^{-a%
\bar{z}/c^{2}}$. The proper frequency of the oscillator $\omega _{0}$ is
given by $\omega _{0}=\omega e^{-a\bar{z}/c^{2}}$. An infinitely long chain
in the Rindler space ($-\infty <\bar{z}<\infty $) covers a semi-infinite
segment $z\geq c|t|$ in Minkowski space. In a realistic situation the chain
has a finite length as shown in Fig. \ref{Fig1}. In this case the speed of
oscillators in the chain is always smaller than $c$ and the linear density
in the Minkowski spacetime is finite everywhere. If the chain length is very
large then it can be approximately modeled as an infinite, which we assume
in the present discussion.

Physically, the Minkowski vacuum is the ground state of free field in
Minkowski spacetime. In the Minkowski vacuum there are no plane-wave
photons. In this Section we will describe the field in terms of Rindler
photons that are present in the Minkowski vacuum. For simplicity, we
consider a scalar field in 1+1 dimension. The right-moving Rindler modes
read \cite{Rind66}%
\begin{equation}
\phi _{R1\Omega }(t,z)=\frac{1}{\sqrt{\Omega }}\left( \frac{z}{c}-t\right)
^{i\Omega }\theta \left( \frac{z}{c}-t\right) ,  \label{R1}
\end{equation}%
\begin{equation}
\phi _{R2\Omega }(t,z)=\frac{1}{\sqrt{\Omega }}\left( t-\frac{z}{c}\right)
^{-i\Omega }\theta \left( t-\frac{z}{c}\right) ,  \label{R2}
\end{equation}%
where $\Omega >0$ is a \textquotedblleft Rindler\textquotedblright\
frequency and $\theta (x)$ is the Heaviside step function. In Eqs. (\ref{R1}%
) and (\ref{R2}), subscripts 1 and 2 refer to the right and left Rindler
wedges, while R and L refer to right and left moving modes (functions of $%
t-z/c$ and $t+z/c$).

The left-moving Rindler modes $\phi _{L1\Omega }$ and $\phi _{L1\Omega }$
are obtained from Eqs. (\ref{R1}) and (\ref{R2}) by coordinate inversion $%
z\rightarrow -z$. Rindler modes $\phi _{R1\Omega }$ ($\phi _{R2\Omega }$)
are nonzero below (above) the line $t=z/c$ (see Fig. \ref{Fig1}). The
corresponding annihilation operators of the Rindler photons we denote as $%
\hat{b}_{R1\Omega }$ and $\hat{b}_{R2\Omega }$ for the right-moving modes,
and $\hat{b}_{L1\Omega }$ and $\hat{b}_{L2\Omega }$ for the left-moving
modes.

Next we introduce operators describing collective excitation of the
oscillators in the chain%
\begin{equation}
\hat{\sigma}_{k}=\frac{1}{\mathcal{N}}\sum_{m=-\infty }^{\infty }e^{-ik\bar{z%
}_{m}}\hat{\sigma}_{m},  \label{CA1}
\end{equation}%
where $\hat{\sigma}_{m}$ is the annihilation operator for the oscillator $m$
located at the point $\bar{z}_{m}$ and $\mathcal{N}$ is a normalization
factor. Since oscillators are moving in the right Rindler wedge (see Fig. %
\ref{Fig1}), they are coupled only with the modes $\phi _{R1\Omega }$ and $%
\phi _{L2\Omega }$ which are nonzero in this wedge. In Appendix \ref{CC1} we
show that the right-moving (left-moving) Rindler mode $\phi _{R1\Omega }$ ($%
\phi _{L2\Omega }$) is coupled with the collective oscillator excitation $%
\hat{\sigma}_{k_{\Omega }}$ ($\hat{\sigma}_{-k_{\Omega }}$), where $%
k_{\Omega }=\Omega a/c^{2}$ and the interaction Hamiltonian between
oscillators and the field in the collective basis reads%
\begin{equation*}
\hat{V}(\tau )\propto \int_{0}^{\infty }d\Omega \sqrt{\Omega }\left( \hat{%
\sigma}_{k_{\Omega }}^{\dag }\hat{b}_{R1\Omega }e^{-i\tau \left( \frac{%
\Omega a}{c}-\omega \right) }\right.
\end{equation*}%
\begin{equation}
\left. +\hat{\sigma}_{-k_{\Omega }}^{\dag }\hat{b}_{L2\Omega }e^{-i\tau
\left( \frac{\Omega a}{c}-\omega \right) }+h.c.\right) .  \label{IG}
\end{equation}

Thus, in terms of the collective chain modes $\hat{\sigma}_{k}$ the
evolution of the system is factorized. Namely, the coupled modes $\hat{\sigma%
}_{k_{\Omega }}$ ($\hat{\sigma}_{-k_{\Omega }}$) and $\hat{b}_{R\Omega }$ ($%
\hat{b}_{L\Omega }$) evolve independently from the rest of the system. As a
result, we can consider evolution of the system for each pair of modes $\hat{%
\sigma}_{k_{\Omega }}$ ($\hat{\sigma}_{-k_{\Omega }}$) and $\hat{b}_{R\Omega
}$ ($\hat{b}_{L\Omega }$) separately. To be specific, we select a
right-moving Rindler mode for which the resonance condition%
\begin{equation*}
\omega =\frac{a\Omega }{c}
\end{equation*}%
is satisfied. This mode of the field is coupled with the collective
oscillator mode $k=\omega /c$ through the interaction Hamiltonian 
\begin{equation}
\hat{V}=\hslash g\left( \hat{\sigma}^{\dag }\hat{b}_{1}+\hat{\sigma}\hat{b}%
_{1}^{\dag }\right) ,  \label{h1b}
\end{equation}%
where $g$ is an effective coupling constant and to simplify notations we use 
$\hat{\sigma}$ for $\hat{\sigma}_{k_{\Omega }}$, and $\hat{b}_{1}$ for $\hat{%
b}_{R1\Omega }$. In Appendix \ref{KK1} we discuss a general case when mode
of the field can be detuned from the resonance.

The Minkowski vacuum $\left\vert 0_{M}\right\rangle $ is filled with Rindler
photons, namely $\left\vert 0_{M}\right\rangle $ is a product of a two-mode
squeezed states for each pair of Rindler photons $\hat{b}_{R1\Omega }$ ($%
\hat{b}_{L1\Omega }$) and $\hat{b}_{R2\Omega }$ ($\hat{b}_{L2\Omega }$). For
the present discussion only state of the photon pair $\hat{b}_{R1\Omega }$
and $\hat{b}_{R2\Omega }$ is relevant, which reads \cite{Unru84} 
\begin{equation}
\left\vert 0_{M}\right\rangle =\sqrt{1-\gamma ^{2}}e^{\gamma \hat{b}%
_{1}^{\dagger }\hat{b}_{2}^{\dagger }}\left\vert 0_{R}\right\rangle =\sqrt{%
1-\gamma ^{2}}\sum_{n=0}^{\infty }\gamma ^{n}\left\vert nn\right\rangle ,
\label{w0}
\end{equation}%
where%
\begin{equation}
\gamma =e^{-\pi \Omega },
\end{equation}%
$\left\vert 0_{R}\right\rangle $ stands for the Rindler vacuum and $%
\left\vert nn\right\rangle $ is a state with $n$ photons in the Rindler
modes $\phi _{1}$ and $\phi _{2}$. Equation (\ref{w0}) shows that the
Minkowski vacuum is entangled when expressed in terms of these Rindler
modes. Namely, the numbers of Rindler photons in the modes $\phi _{1}$ and $%
\phi _{2}$ are correlated. For example, if there are $n$ photons in the mode 
$\phi _{1}$, then with unit probability there are $n$ photons in the mode $%
\phi _{2}$.

Collective mode of the oscillator chain $\hat{\sigma}$ can become excited by
absorbing Rindler photons $\hat{b}_{1}$ which are present in the Minkowski
vacuum. The Schrodinger equation for the evolution of the system's state
vector%
\begin{equation*}
i\hslash \frac{\partial }{\partial \tau }\left\vert \psi (\tau
)\right\rangle =\hat{V}\left\vert \psi (\tau )\right\rangle ,
\end{equation*}%
with $\hat{V}$ given by Eq. (\ref{h1b}), yields%
\begin{equation}
\left\vert \psi (\tau )\right\rangle =e^{-ig\tau \left( \hat{\sigma}^{\dag }%
\hat{b}_{1}+\hat{\sigma}\hat{b}_{1}^{\dag }\right) }\left\vert
G\right\rangle \left\vert 0_{M}\right\rangle ,  \label{q0bb}
\end{equation}%
where $\left\vert 0_{M}\right\rangle $ is the initial state vector of the
field (Minkowski vacuum), and $\left\vert G\right\rangle $ is the initial
state of the oscillator chain which we assume to be the ground state. State
vector (\ref{q0bb}) can be written as (see Appendix \ref{AA1})%
\begin{equation}
\left\vert \psi (\tau )\right\rangle =e^{\gamma \hat{b}_{2}^{\dagger }\left(
\left( \cos (g\tau )-1\right) \hat{b}_{1}^{\dag }-i\sin (g\tau )\hat{\sigma}%
^{\dag }\right) }\left\vert G\right\rangle \left\vert 0_{M}\right\rangle ,
\label{xx1}
\end{equation}%
or in terms of the Rindler vacuum $\left\vert 0_{R}\right\rangle $%
\begin{equation}
\left\vert \psi (\tau )\right\rangle =\sqrt{1-\gamma ^{2}}e^{\gamma \hat{b}%
_{2}^{\dagger }\left( \cos (g\tau )\hat{b}_{1}^{\dag }-i\sin (g\tau )\hat{%
\sigma}^{\dag }\right) }\left\vert G\right\rangle \left\vert
0_{R}\right\rangle .  \label{m7c}
\end{equation}%
Equation (\ref{m7c}) shows that state of the system periodically oscillates
with $\tau $ with the Rabi frequency $g$. At the time instances for which $%
\cos (g\tau )=0$ the state reduces to%
\begin{equation*}
\left\vert \psi \right\rangle =\sqrt{1-\gamma ^{2}}e^{\pm i\gamma \hat{\sigma%
}^{\dag }\hat{b}_{2}^{\dagger }}\left\vert G\right\rangle \left\vert
0_{R}\right\rangle ,
\end{equation*}%
which is a two-mode squeezed state between the oscillators and the Rindler
photons $\hat{b}_{2}$, and no photons are present in the mode $\hat{b}_{1}$.
That is the initial entanglement between Rindler photons $\hat{b}_{1}$ and $%
\hat{b}_{2}$ is transferred to the entanglement between the oscillators and
photons $\hat{b}_{2}$.

At the time instances for which $\cos (g\tau )=-1$ the state of the system
becomes%
\begin{equation*}
\left\vert \psi \right\rangle =\sqrt{1-\gamma ^{2}}e^{-\gamma \hat{b}%
_{1}^{\dagger }\hat{b}_{2}^{\dagger }}\left\vert G\right\rangle \left\vert
0_{R}\right\rangle .
\end{equation*}%
That is oscillator chain is back to the ground state, but state of the field
differs from the original Minkowski vacuum (\ref{w0}) by change $\gamma
\rightarrow -\gamma $.

One can obtain an average number of collective oscillator excitations and
photons in the mode $\hat{b}_{1}$ directly from the Hamiltonian (\ref{h1b})
using the Heisenberg picture in which operators $\hat{b}_{1}(\tau )$ and $%
\hat{\sigma}(\tau )$ obey equation of motion%
\begin{equation*}
i\hslash \frac{\partial \hat{\sigma}(\tau )}{\partial \tau }=\left[ \hat{%
\sigma}(\tau ),\hat{V}\right] =\hslash g\hat{b}_{1}(\tau ),
\end{equation*}%
\begin{equation*}
i\hslash \frac{\partial \hat{b}_{1}(\tau )}{\partial \tau }=\left[ \hat{b}%
_{1}(\tau ),\hat{V}\right] =\hslash g\hat{\sigma}(\tau ).
\end{equation*}%
Solution of these equations is%
\begin{equation*}
\hat{\sigma}(\tau )=\cos \left( g\tau \right) \hat{\sigma}-i\sin \left(
g\tau \right) \hat{b}_{1},
\end{equation*}%
\begin{equation*}
\hat{b}_{1}(\tau )=\cos \left( g\tau \right) \hat{b}_{1}-i\sin \left( g\tau
\right) \hat{\sigma}.
\end{equation*}%
Taking the expectation value over the initial state vector of the system, we
find that the average number of the chain collective excitations $N_{\sigma
} $ undergoes Rabi oscillations%
\begin{equation*}
N_{\sigma }(\tau )=\frac{\gamma ^{2}}{1-\gamma ^{2}}\sin ^{2}\left( g\tau
\right) ,
\end{equation*}%
together with the average number of Rindler photons $\hat{b}_{1}$ 
\begin{equation*}
N_{b_{1}}(\tau )=\frac{\gamma ^{2}}{1-\gamma ^{2}}\cos ^{2}\left( g\tau
\right) .
\end{equation*}%
The Rabi oscillation dynamics can be also obtained by treating the field and
the oscillator chain classically (see Appendix \ref{KK1}).

Using Eq. (\ref{m7c}) one can calculate the reduced density operator for the
photons in the mode $\hat{b}_{1}$. We find (see Appendix \ref{AA2})%
\begin{equation}
\hat{\rho}_{b_{1}}(\tau )=\left( 1-\gamma ^{2}\right) \sum_{m=0}^{\infty }%
\frac{\gamma ^{2m}\cos ^{2m}(g\tau )}{\left( 1-\gamma ^{2}\sin ^{2}(g\tau
)\right) ^{m+1}}\left\vert m\right\rangle \left\langle m\right\vert ,
\label{w2b}
\end{equation}%
where $\left\vert m\right\rangle $ is a state with $m$ photons in the mode $%
\hat{b}_{1}$. Equation (\ref{w2b}) shows that statistics of photons in the
mode $\hat{b}_{1}$ is thermal at any $\tau $. Equation (\ref{w2b}) can be
written as 
\begin{equation*}
\hat{\rho}_{b_{1}}(\tau )=\frac{1-\gamma ^{2}}{1-\gamma ^{2}\sin ^{2}(g\tau )%
}e^{\ln \left( \frac{\gamma ^{2}\cos ^{2}(g\tau )}{1-\gamma ^{2}\sin
^{2}(g\tau )}\right) \hat{b}_{1}^{\dag }\hat{b}_{1}}.
\end{equation*}

Similarly, we obtain the reduced density operator for the oscillator
collective mode%
\begin{equation*}
\hat{\rho}_{\sigma }(\tau )=\frac{1-\gamma ^{2}}{1-\gamma ^{2}\cos
^{2}(g\tau )}e^{\ln \left( \frac{\gamma ^{2}\sin ^{2}(g\tau )}{1-\gamma
^{2}\cos ^{2}(g\tau )}\right) \hat{\sigma}^{\dag }\hat{\sigma}}.
\end{equation*}%
Statistics of the collective oscillator excitations is also thermal at any $%
\tau $.

\section{Emission of entangled Unruh-Minkowski photons by accelerated
oscillator chain}

If we describe the field by means of the right-moving Unruh-Minkowski
photons $\hat{a}_{1}$ and $\hat{a}_{2}$ with mode functions 
\begin{equation}
F_{1}(t,z)=\frac{\phi _{1}(t,z)-\gamma \phi _{2}^{\ast }(t,z)}{\sqrt{%
1-\gamma ^{2}}},  \label{UM}
\end{equation}%
and%
\begin{equation}
F_{2}(t,z)=\frac{\phi _{2}(t,z)-\gamma \phi _{1}^{\ast }(t,z)}{\sqrt{%
1-\gamma ^{2}}},  \label{UM2}
\end{equation}%
where $\phi _{1}$ and $\phi _{2}$ are the right-moving Rindler modes and $%
\gamma =e^{-\pi \Omega }$, the Minkowski vacuum looks empty \cite{Unru84}.
In this description, the accelerated oscillator becomes excited by emitting
Unruh-Minkowski photon $\hat{a}_{2}$ which, from the oscillator's
perspective, has negative frequency \cite{Svid21}. The latter guarantees
conservation of energy during oscillator excitation. We call this the
Unruh-Wald effect \cite{Unru84}.

One can rewrite the state vector of the system (\ref{m7c}) in terms of the
Unruh-Minkowski photons operators $\hat{a}_{1}$ and $\hat{a}_{2}$ using the
following relations between operators for the Rindler photons and for the
Unruh-Minkowski photons \cite{Unru84}

\begin{equation}
\hat{b}_{1}=\frac{\hat{a}_{1}+\gamma \hat{a}_{2}^{\dagger }}{\sqrt{1-\gamma
^{2}}},\quad \hat{b}_{1}^{\dagger }=\frac{\hat{a}_{1}^{\dagger }+\gamma \hat{%
a}_{2}}{\sqrt{1-\gamma ^{2}}},  \label{f3a}
\end{equation}%
\begin{equation}
\hat{b}_{2}=\frac{\hat{a}_{2}+\gamma \hat{a}_{1}^{\dagger }}{\sqrt{1-\gamma
^{2}}},\quad \hat{b}_{2}^{\dagger }=\frac{\hat{a}_{2}^{\dagger }+\gamma \hat{%
a}_{1}}{\sqrt{1-\gamma ^{2}}}.  \label{f4a}
\end{equation}%
Details of the calculations are provided in Appendix \ref{AA3}. We find%
\begin{equation*}
\left\vert \psi (\tau )\right\rangle =\frac{1-\gamma ^{2}}{1-\gamma ^{2}\cos
(g\tau )}e^{-i\gamma \sqrt{1-\gamma ^{2}}\frac{\sin (g\tau )}{1-\gamma
^{2}\cos (g\tau )}\hat{a}_{2}^{\dagger }\hat{\sigma}^{\dag }}
\end{equation*}%
\begin{equation}
\times e^{-\frac{\gamma \left( 1-\cos (g\tau )\right) }{1-\gamma ^{2}\cos
(g\tau )}\hat{a}_{1}^{\dagger }\hat{a}_{2}^{\dagger }}\left\vert
G\right\rangle \left\vert 0_{M}\right\rangle .  \label{x5}
\end{equation}%
That is an accelerated oscillator chain generates photons $\hat{a}_{1}$ and $%
\hat{a}_{2}$. For the time instances for which $\cos (g\tau )=-1$, the state
vector of the system reduces to 
\begin{equation}
\left\vert \psi \right\rangle =\frac{1-\gamma ^{2}}{1+\gamma ^{2}}e^{-\frac{%
2\gamma }{1+\gamma ^{2}}\hat{a}_{1}^{\dagger }\hat{a}_{2}^{\dagger
}}\left\vert G\right\rangle \left\vert 0_{M}\right\rangle ,
\end{equation}%
which is a two-mode squeezed state of the Unruh-Minkowski photons $\hat{a}%
_{1}$ and $\hat{a}_{2}$, and the oscillator chain is in the ground state $%
\left\vert G\right\rangle $. Thus, accelerated oscillator chain emits
entangled pairs of photons $\hat{a}_{1}$ and $\hat{a}_{2}$.

Generation of the entangled Unruh-Minkowski photon pairs can be understood
as follows \cite{Scul22}. First, the ground-state oscillator becomes excited
by emitting photon $\hat{a}_{2}$ which, from the oscillator's perspective,
has negative frequency (energy) \cite{Svid21}. This process is followed by
spontaneous decay of the oscillator back to the ground state with emission
of photon $\hat{a}_{1}$ with positive frequency.

\section{Two uniformly accelerated oscillator chains in Minkowski vacuum}

\begin{figure}[h]
\begin{center}
\epsfig{figure=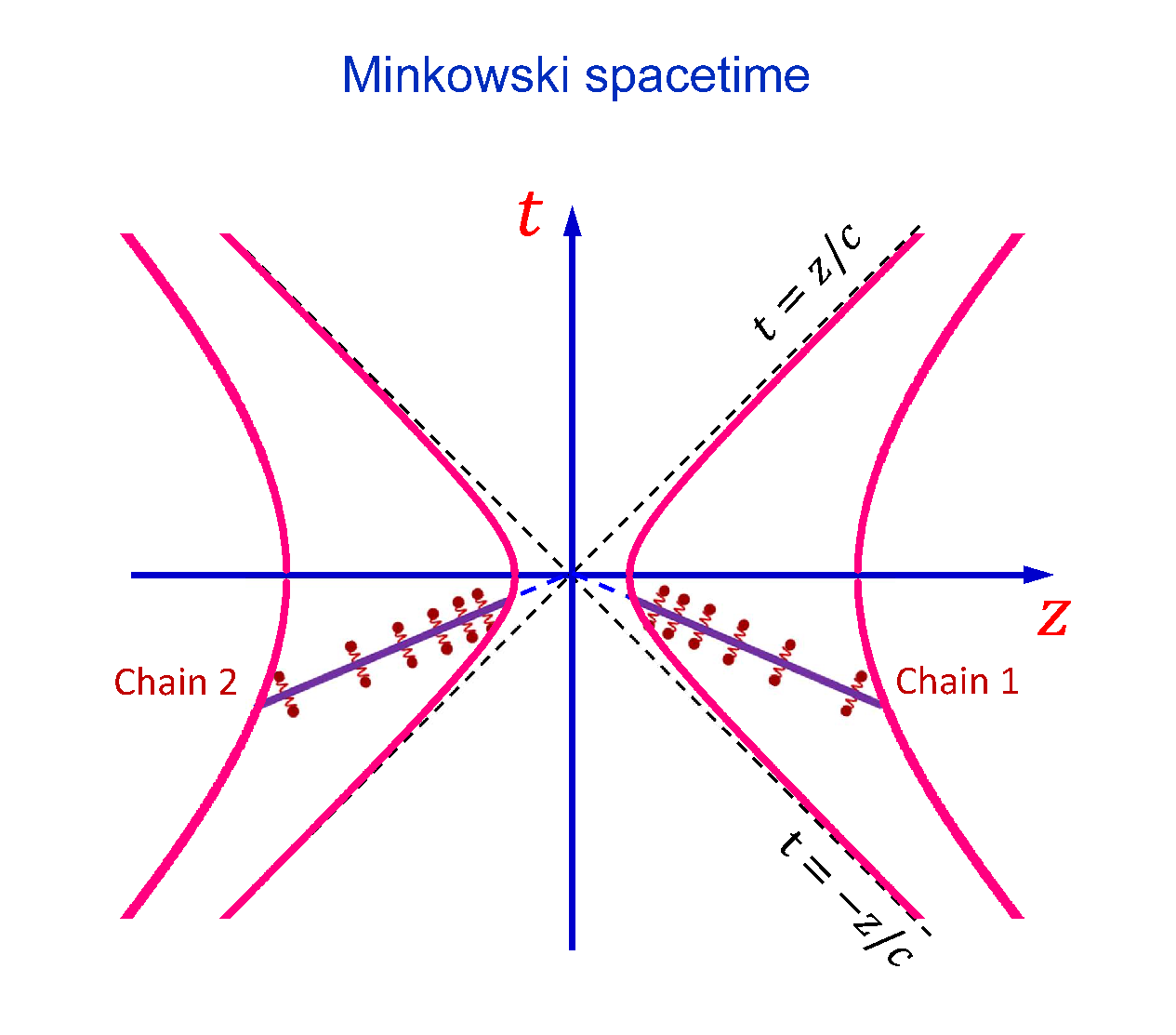, angle=0, width=8cm}
\end{center}
\caption{Two identical harmonic oscillator chains are accelerated in the
opposite Rindler wedges through Minkowski vacuum.}
\label{Fig2}
\end{figure}

Entanglement of the Minkowski vacuum yields correlated excitation of atoms
accelerated in causally disconnected regions \cite{Svid21a}. Here we
investigate dynamics of this process and consider two identical harmonic
oscillator chains accelerated with the same $|a|$ in the right and left
Rindler wedges respectively through the Minkowski vacuum $\left\vert
0_{M}\right\rangle $ (see Fig. \ref{Fig2}). Trajectories of the oscillators
in the chains are%
\begin{equation*}
t_{1}=\frac{c}{a}e^{a\bar{z}/c^{2}}\sinh \left( \frac{a\tau }{c}\right)
,\quad z_{1}=\frac{c^{2}}{a}e^{a\bar{z}/c^{2}}\cosh \left( \frac{a\tau }{c}%
\right) ,
\end{equation*}%
and%
\begin{equation*}
t_{2}=\frac{c}{a}e^{-a\bar{z}/c^{2}}\sinh \left( \frac{a\tau }{c}\right)
,\quad z_{2}=-\frac{c^{2}}{a}e^{-a\bar{z}/c^{2}}\cosh \left( \frac{a\tau }{c}%
\right) ,
\end{equation*}%
where $a>0$ and $\tau $ is the Rindler time. The chains move in causally
disconnected regions and do not affect each other. However, they interact
with the same field and, as we show below, the oscillator chains become
excited in correlated fashion.

As before, we will consider the problem in 1+1 dimensions and consider the
right-moving Rindler modes $\hat{b}_{R1\Omega }$ and $\hat{b}_{R2\Omega }$
that are coupled with the collective chain modes $\hat{\sigma}_{1k_{\Omega
}} $ and $\hat{\sigma}_{2k_{\Omega }}$, where $k_{\Omega }=\Omega a/c^{2}$,
respectively. If in the Rindler time all oscillators have the same frequency 
$\omega =a\Omega /c$ the corresponding interaction Hamiltonian reads%
\begin{equation}
\hat{V}=\hslash g\left( \hat{\sigma}_{1}^{\dag }\hat{b}_{1}+\hat{\sigma}_{1}%
\hat{b}_{1}^{\dag }+\hat{\sigma}_{2}^{\dag }\hat{b}_{2}+\hat{\sigma}_{2}\hat{%
b}_{2}^{\dag }\right) ,  \label{h1}
\end{equation}%
where to simplify notations we introduced $\hat{\sigma}_{1}=\hat{\sigma}%
_{1k_{\Omega }}$, $\hat{\sigma}_{2}=\hat{\sigma}_{2k_{\Omega }}$, $\hat{b}%
_{1}=\hat{b}_{R1\Omega }$ and $\hat{b}_{2}=\hat{b}_{R2\Omega }$.

Schrodinger equation for the evolution of the system's state vector yields%
\begin{equation}
\left\vert \psi (\tau )\right\rangle =e^{-ig\tau \left( \hat{\sigma}%
_{1}^{\dag }\hat{b}_{1}+\hat{\sigma}_{1}\hat{b}_{1}^{\dag }+\hat{\sigma}%
_{2}^{\dag }\hat{b}_{2}+\hat{\sigma}_{2}\hat{b}_{2}^{\dag }\right)
}\left\vert G\right\rangle \left\vert 0_{M}\right\rangle ,  \label{q0}
\end{equation}%
where $\left\vert 0_{M}\right\rangle $ is the initial state vector of the
field, and $\left\vert G\right\rangle $ is the initial state of the
oscillators which we assume to be the ground state. Equation (\ref{q0}) can
be written as (see Appendix \ref{AA4})%
\begin{equation*}
\left\vert \psi (\tau )\right\rangle =\exp \left[ \frac{\gamma }{2}\left(
\left( \cos (2g\tau )-1\right) \left( \hat{b}_{1}^{\dag }\hat{b}%
_{2}^{\dagger }+\hat{\sigma}_{1}^{\dag }\hat{\sigma}_{2}^{\dag }\right)
\right. \right.
\end{equation*}%
\begin{equation}
\left. \left. -i\sin (2g\tau )\left( \hat{b}_{2}^{\dag }\hat{\sigma}%
_{1}^{\dag }+\hat{b}_{1}^{\dag }\hat{\sigma}_{2}^{\dag }\right) \right) %
\right] \left\vert G\right\rangle \left\vert 0_{M}\right\rangle ,  \label{m8}
\end{equation}%
or in terms of the Rindler vacuum%
\begin{equation*}
\left\vert \psi (\tau )\right\rangle =\sqrt{1-\gamma ^{2}}\exp \left[ \gamma
\left( \cos (g\tau )\hat{b}_{1}^{\dag }-i\sin (g\tau )\hat{\sigma}_{1}^{\dag
}\right) \right.
\end{equation*}%
\begin{equation}
\left. \left( \cos (g\tau )\hat{b}_{2}^{\dag }-i\sin (g\tau )\hat{\sigma}%
_{2}^{\dag }\right) \right] \left\vert G\right\rangle \left\vert
0_{R}\right\rangle .  \label{m7}
\end{equation}

Equation (\ref{m7}) shows that state of the system periodically oscillates
with $\tau $ with the Rabi frequency $2g$. At the moments of time for which $%
\cos (g\tau )=0$, the state vector reduces to 
\begin{equation}
\left\vert \psi \right\rangle =\sqrt{1-\gamma ^{2}}e^{-\gamma \hat{\sigma}%
_{1}^{\dag }\hat{\sigma}_{2}^{\dag }}\left\vert G\right\rangle \left\vert
0_{R}\right\rangle ,  \label{m7a}
\end{equation}%
which is a two-mode squeezed state for the two oscillator chains, and the
field is in the Rindler vacuum. That is, initial entanglement of the Rindler
photons in the Minkowski vacuum is transferred to the entanglement of the
oscillator chains moving in the causally disconnected regions. The state of
the field periodically oscillates between the Minkowski vacuum and the
Rindler vacuum $\left\vert 0_{R}\right\rangle $. That is present
configuration allows us to produce Rindler vacuum out of the Minkowski
vacuum.

\section{Duality of oscillator-field entanglement}

In the previous Section we show that ground-state oscillators accelerated in
causally disconnected regions become entangled due to initial entanglement
of the field in terms of Rindler photons. Here we show that this point of
view is relative and one can also think of the process as entanglement
transfer from the oscillators to the field. Namely, collective modes of two
ground-state oscillators are correlated. This correlation is transferred to
the field in the process of photon emission.

To be specific, we consider two systems ($A$ and $B$); each of them is
described by two modes. Operators of excitations in these modes we denote as 
$\hat{A}_{1}$, $\hat{A}_{2}$, $\hat{B}_{1}$, and $\hat{B}_{2}$, and assume
that these operators obey bosonic commutation relations%
\begin{equation}
\lbrack \hat{A}_{1},\hat{A}_{1}^{\dag }]=1,\quad \lbrack \hat{A}_{2},\hat{A}%
_{2}^{\dag }]=1,  \label{c1b}
\end{equation}%
\begin{equation}
\lbrack \hat{B}_{1},\hat{B}_{1}^{\dag }]=1,\quad \lbrack \hat{B}_{2},\hat{B}%
_{2}^{\dag }]=1,  \label{c2b}
\end{equation}%
and all other commutators are equal to zero. We assume that systems $A$ and $%
B$ interact with each other via the Hamiltonian 
\begin{equation}
\hat{V}=\hslash g\left( \hat{A}_{1}^{\dag }\hat{B}_{1}+\hat{A}_{1}\hat{B}%
_{1}^{\dag }+\hat{A}_{2}^{\dag }\hat{B}_{2}+\hat{A}_{2}\hat{B}_{2}^{\dag
}\right) ,  \label{p0}
\end{equation}%
and the initial state $\left\vert \psi _{0}\right\rangle $ is vacuum for the
operators $\hat{A}_{1}$, $\hat{A}_{2}$, and a two-mode squeezed state for
the operators $\hat{B}_{1}$ and $\hat{B}_{2}$%
\begin{equation}
\left\vert \psi _{0}\right\rangle =\sqrt{1-\gamma ^{2}}e^{\gamma \hat{B}%
_{1}^{\dag }\hat{B}_{2}^{\dag }}\left\vert 0_{B}\right\rangle \left\vert
0_{A}\right\rangle .  \label{p0a}
\end{equation}%
Similarly to derivation of Eq. (\ref{m7}), Schrodinger equation for the
evolution of the state vector%
\begin{equation*}
i\hslash \frac{\partial }{\partial \tau }\left\vert \psi (t)\right\rangle =%
\hat{V}\left\vert \psi (\tau )\right\rangle
\end{equation*}%
yields%
\begin{equation*}
\left\vert \psi (\tau )\right\rangle =\sqrt{1-\gamma ^{2}}\exp \left[ \gamma
\left( \cos (g\tau )\hat{B}_{1}^{\dag }-i\sin (g\tau )\hat{A}_{1}^{\dag
}\right) \right.
\end{equation*}%
\begin{equation}
\left. \left( \cos (g\tau )\hat{B}_{2}^{\dag }-i\sin (g\tau )\hat{A}%
_{2}^{\dag }\right) \right] \left\vert 0_{B}\right\rangle \left\vert
0_{A}\right\rangle .  \label{p1}
\end{equation}

That is state of the system periodically oscillates with the Rabi frequency $%
2g$. At the moments of time $\tau $ for which $\cos (g\tau )=0$, the state
vector reduces to%
\begin{equation*}
\left\vert \psi \right\rangle =\sqrt{1-\gamma ^{2}}e^{-\gamma \hat{A}%
_{1}^{\dag }\hat{A}_{2}^{\dag }}\left\vert 0_{B}\right\rangle \left\vert
0_{A}\right\rangle ,
\end{equation*}%
which is a two-mode squeezed state for the modes $\hat{A}_{1}$ and $\hat{A}%
_{2}$, and the field is in the vacuum state for the operators $\hat{B}_{1}$
and $\hat{B}_{2}$. That is, initial entanglement of the $\hat{B}_{1}$ and $%
\hat{B}_{2}$ excitations is transferred to the entanglement of the
excitations in the modes $\hat{A}_{1}$ and $\hat{A}_{2}$.

Next we discuss two uniformly accelerated oscillator chains of the previous
Section from two different perspectives.

\subsection{Description in terms of Rindler photons}

The interaction Hamiltonian between the oscillator chains and Rindler
photons (\ref{h1}) has the same form as Eq. (\ref{p0}) if we identify%
\begin{equation*}
\hat{A}_{1}=\hat{\sigma}_{1},\quad \hat{A}_{2}=\hat{\sigma}_{2},\quad \hat{B}%
_{1}=\hat{b}_{1},\quad \hat{B}_{2}=\hat{b}_{2},
\end{equation*}%
while the initial state of the oscillator-field system%
\begin{equation*}
\left\vert \psi _{0}\right\rangle =\sqrt{1-\gamma ^{2}}e^{\gamma \hat{b}%
_{1}^{\dagger }\hat{b}_{2}^{\dagger }}\left\vert 0_{R}\right\rangle
\left\vert G\right\rangle ,
\end{equation*}%
is the same as the state (\ref{p0a}). Evolution of the system's state vector
as a function of the oscillator's proper time $\tau $ is given by Eq. (\ref%
{m7}). At the moments of time for which $\cos (g\tau )=0$, the state vector
reduces to Eq. (\ref{m7a}) which is a two-mode squeezed state for the two
oscillator chains, and the field is in the Rindler vacuum $\left\vert
0_{R}\right\rangle $. This result is naturally interpreted as if initial
entanglement of Rindler photons in the Minkowski vacuum is transferred to
the entanglement of the oscillators moving in the causally disconnected
regions.

\subsection{Description in terms of Unruh-Minkowski photons}

In terms of Unruh-Minkowski photons the initial state of the field has no
particles, and initially both oscillator chains are in the ground state.
Motion of the oscillator chains in the causally disconnected regions yields
generation of entangled pairs of Unruh-Minkowski photons $\hat{a}_{1}$ and $%
\hat{a}_{2}$, and during this process the oscillator chains also become
entangled with each other. What is the source of this entanglement?

To answer this question we recall that operators for the Rindler photons and
for the Unruh-Minkowski photons are related by the Bogoliubov
transformations (\ref{f3a}) and (\ref{f4a}). Using these relations, one can
rewrite the interaction Hamiltonian (\ref{h1}) as%
\begin{equation*}
\hat{V}=\frac{\hslash g}{\sqrt{1-\gamma ^{2}}}\left( \hat{a}_{1}^{\dagger
}\left( \hat{\sigma}_{1}+\gamma \hat{\sigma}_{2}^{\dag }\right) +\hat{a}%
_{1}\left( \hat{\sigma}_{1}^{\dag }+\gamma \hat{\sigma}_{2}\right) \right.
\end{equation*}%
\begin{equation*}
+\left. \hat{a}_{2}^{\dagger }\left( \hat{\sigma}_{2}+\gamma \hat{\sigma}%
_{1}^{\dag }\right) +\hat{a}_{2}\left( \hat{\sigma}_{2}^{\dag }+\gamma \hat{%
\sigma}_{1}\right) \right) .
\end{equation*}%
Introducing operators%
\begin{equation*}
\hat{A}_{1}=\hat{a}_{1},\quad \hat{A}_{2}=\hat{a}_{2},
\end{equation*}%
\begin{equation*}
\hat{B}_{1}=\frac{\hat{\sigma}_{1}+\gamma \hat{\sigma}_{2}^{\dag }}{\sqrt{%
1-\gamma ^{2}}},\quad \hat{B}_{2}=\frac{\hat{\sigma}_{2}+\gamma \hat{\sigma}%
_{1}^{\dag }}{\sqrt{1-\gamma ^{2}}},
\end{equation*}%
which obey bosonic commutation relations (\ref{c1b}) and (\ref{c2b}), we have%
\begin{equation*}
\hat{V}=\hslash g\left( \hat{A}_{1}^{\dag }\hat{B}_{1}+\hat{A}_{1}\hat{B}%
_{1}^{\dag }+\hat{A}_{2}^{\dag }\hat{B}_{2}+\hat{A}_{2}\hat{B}_{2}^{\dag
}\right) .
\end{equation*}%
In terms of these operators, the initial state of the system is given by Eq.
(\ref{p0a}). That is, the ground state of two oscillator chains $\left\vert
G\right\rangle $ in terms of the operators $\hat{B}_{1}$ and $\hat{B}_{2}$
is a two-mode squeezed state%
\begin{equation*}
\left\vert G\right\rangle =\sqrt{1-\gamma ^{2}}e^{\gamma \hat{B}_{1}^{\dag }%
\hat{B}_{2}^{\dag }}\left\vert 0_{B}\right\rangle ,
\end{equation*}%
where $\left\vert 0_{B}\right\rangle $ is a vacuum state for $\hat{B}_{1}$
and $\hat{B}_{2}$. Thus, initial state is filled with excitations $\hat{B}%
_{1}$ and $\hat{B}_{2}$ and the latter are entangled. This shows that two
ground-state oscillator chains are entangled in the collective basis $\hat{B}%
_{1}$ and $\hat{B}_{2}$.

Equation (\ref{p1}) yields for the evolution of the state vector%
\begin{equation*}
\left\vert \psi (\tau )\right\rangle =\sqrt{1-\gamma ^{2}}\exp \left[ \gamma
\left( \cos (g\tau )\hat{B}_{1}^{\dag }-i\sin (g\tau )\hat{a}_{1}^{\dag
}\right) \right.
\end{equation*}%
\begin{equation*}
\left. \left( \cos (g\tau )\hat{B}_{2}^{\dag }-i\sin (g\tau )\hat{a}%
_{2}^{\dag }\right) \right] \left\vert 0_{M}\right\rangle \left\vert
0_{B}\right\rangle .
\end{equation*}%
At the moments of time for which $\cos (g\tau )=0$, the state vector reduces
to%
\begin{equation*}
\left\vert \psi \right\rangle =\sqrt{1-\gamma ^{2}}e^{-\gamma \hat{a}%
_{1}^{\dag }\hat{a}_{2}^{\dag }}\left\vert 0_{M}\right\rangle \left\vert
0_{B}\right\rangle ,
\end{equation*}%
which is a two-mode squeezed state for the Unruh-Minkowski photons $\hat{a}%
_{1}$ and $\hat{a}_{2}$, while the oscillators are in the vacuum state for
the operators $\hat{B}_{1}$ and $\hat{B}_{2}$. The latter can be expressed
in terms of $\left\vert G\right\rangle $ as%
\begin{equation*}
\left\vert 0_{B}\right\rangle =\sqrt{1-\gamma ^{2}}e^{-\gamma \hat{\sigma}%
_{1}^{\dag }\hat{\sigma}_{2}^{\dag }}\left\vert G\right\rangle .
\end{equation*}

One can interpret this result as if initial entanglement of the excitations $%
\hat{B}_{1}$ and $\hat{B}_{2}$ is transferred to the entanglement of the
generated Unruh-Minkowski photons $\hat{a}_{1}$ and $\hat{a}_{2}$, and
entanglement between two oscillators in the bare basis $\hat{\sigma}_{1}$
and $\hat{\sigma}_{2}$.

\section{Summary}

Particle content of a field state depends on the mode functions we adopt to
describe photons. If we choose plane-wave modes to quantize the field, the
Minkowski vacuum has no particles. However, if we describe the field in
terms of Rindler photons, the Minkowski vacuum is filled with particles in a
squeezed state and the number of Rindler photons in the modes $\phi _{1}$
and $\phi _{2}$ is correlated, that is vacuum is entangled. Such
entanglement can be transferred to atoms interacting with the field, even if
the atoms are causally disconnected.

Uniformly accelerated atoms interact resonantly with Rindler photons, that
is, in the atom's frame, the mode functions of Rindler photons harmonically
oscillate as a function of the atom's proper time \cite{Svid21}. This is
analogous to the resonant interaction between plane-wave photons and atoms
moving with a constant velocity. Thus, a uniformly accelerated atom can
become excited by resonant absorption of a Rindler photon present in the
Minkowski vacuum, which is known as the Unruh effect \cite%
{Full73,Davi75,Unru76}.

An atom uniformly accelerated in the right Rindler wedge (see Fig. \ref{Fig1}%
) interacts only with the Rindler photons $\phi _{1}$, but not $\phi _{2}$,
since the mode function $\phi _{2}$ vanishes along the atom's worldline.
Tracing over the mode $\phi _{2}$ leaves the remaining mode $\phi _{1}$ in a
thermal state with Unruh temperature. Thus, uniformly accelerated atom feels
a thermal bath of photons $\phi _{1}$ which leads to atom's thermal
excitation. However, due to the initial correlations between photon numbers
in the modes $\phi _{1}$ and $\phi _{2}$ the excited atom becomes entangled
with photons $\phi _{2}$, even though it does not interact with them.

Here we investigated this process for a long harmonic oscillator chain
accelerated in the right Rindler wedge. In this case the system's evolution
can be factorized. Namely, there are collective excitations of the
oscillators which are coupled with only one Rindler mode of the field. We
find an exact analytical solution for the state vector of these coupled
modes which undergoes periodic Rabi oscillations as a function of the
oscillator's proper time $\tau $. We find that for the resonant mode $\phi
_{1}$ after half of the Rabi cycle, the oscillator chain becomes maximally
excited by absorbing all Rindler photons from the mode $\phi _{1}$ leaving
mode $\phi _{1}$ in the state of Rindler vacuum. In such moment of time the
excited oscillator chain is in the two-mode squeezed state with photons $%
\phi _{2}$.

If we describe the field by means of the Unruh-Minkowski photons $\hat{a}%
_{1} $ and $\hat{a}_{2}$ the Minkowski vacuum looks empty. In this
description, a uniformly accelerated oscillator becomes excited by emitting
Unruh-Minkowski photon $\hat{a}_{2}$ (we call this the Unruh-Wald effect 
\cite{Unru84}) which, from the oscillator's perspective, has negative
frequency \cite{Svid21}. The latter guarantees conservation of energy during
oscillator excitation. This process is followed by a rapid spontaneous decay
of the oscillator back to the ground state with emission of a photon $\hat{a}%
_{1}$ with positive frequency leading to the generation of entangled
Unruh-Minkowski photon pairs ($\hat{a}_{1}$ and $\hat{a}_{2}$) in a two-mode
squeezed state \cite{Scul22}. We find an exact analytical solution for the
system's state vector as a function of $\tau $ for the oscillator chain
accelerated through the Minkowski vacuum. The obtained solution shows that
the system undergoes Rabi oscillations describing photon emission and
re-absorption.

Entanglement of the Minkowski vacuum can lead to correlated excitation of
two causally disconnected oscillator chains accelerated in the opposite
Rindler wedges (see Fig. \ref{Fig2}). That is, initial entanglement of the
Rindler photons present in the Minkowski vacuum is transferred to the
entanglement of the oscillators interacting with the same field. We show
that for the resonant modes the state of the field periodically oscillates
between the Minkowski vacuum and the Rindler vacuum for modes $\phi _{1}$
and $\phi _{2}$. That is such configuration, apart from entanglement
harvesting, allows us to produce Rindler vacuum out of the Minkowski vacuum.

In terms of Unruh-Minkowski photons the initial state of the field has no
particles. We find that motion of the ground-state oscillator chains in the
causally disconnected regions yields generation of entangled pairs of
Unruh-Minkowski photons $\hat{a}_{1}$ and $\hat{a}_{2}$. During this process
the oscillator chains also become excited and entangled with each other. We
show that one can interpret this result as if initial entanglement between
the collective oscillator excitations $\hat{B}_{1}$ and $\hat{B}_{2}$ is
transferred to the entanglement of the generated Unruh-Minkowski photons $%
\hat{a}_{1}$ and $\hat{a}_{2}$, and entanglement between the two chains.
Such ground-state correlations are an inherent property of many-body systems.

\begin{acknowledgements}
This work was supported by U.S. Department of Energy (DE-SC-0023103, FWP-ERW7011, DE-SC0024882); Welch Foundation (A-1261); 
National Science Foundation (PHY-2013771); Air Force Office of Scientific Research (FA9550-20-1-0366).
W.U. thanks the Natural Sciences and Engineering Research Council of Canada (NSERC) (Grant No. 5-80441),
and the TAMU Hagler Institute for Advanced Studies for their support.
\end{acknowledgements}

\appendix


\section{Accelerated chain of oscillators}

\label{CC1}

An accelerated point-like oscillator interacts with many Rindler modes. Here
we consider a system of oscillators whose collective excitations interact
with only one Rindler mode and, hence, the system's evolution can be
factorized.

Rindler modes reduce to plane-waves in Rindler space. The coordinate
transformation from the Minkowski space $(t,z)$ into the Rindler space $%
(\tau ,\bar{z})$ is%
\begin{equation}
t(\tau ,\bar{z})=\frac{c}{a}e^{a\bar{z}/c^{2}}\sinh \left( \frac{a\tau }{c}%
\right) ,  \label{RS1}
\end{equation}%
\begin{equation}
z(\tau ,\bar{z})=\frac{c^{2}}{a}e^{a\bar{z}/c^{2}}\cosh \left( \frac{a\tau }{%
c}\right) ,  \label{RS2}
\end{equation}%
where $a>0$ is a parameter with dimension of acceleration. In Rindler space
the scalar field operator $\hat{\Phi}(\tau ,\bar{z})$ in terms of the right $%
R$ and left $L$ propagating Rindler modes reads%
\begin{equation*}
\hat{\Phi}(\tau ,\bar{z})=i\int_{0}^{\infty }\frac{d\Omega }{\sqrt{4\pi
\Omega }}\left( \hat{b}_{R\Omega }e^{-i\frac{\Omega a}{c}\left( \tau -\frac{%
\bar{z}}{c}\right) }-\hat{b}_{R\Omega }^{\dagger }e^{i\frac{\Omega a}{c}%
\left( \tau -\frac{\bar{z}}{c}\right) }\right.
\end{equation*}%
\begin{equation*}
\left. +\hat{b}_{L\Omega }e^{-i\frac{\Omega a}{c}\left( \tau +\frac{\bar{z}}{%
c}\right) }-\hat{b}_{L\Omega }^{\dagger }e^{i\frac{\Omega a}{c}\left( \tau +%
\frac{\bar{z}}{c}\right) }\right) ,
\end{equation*}%
where $\hat{b}_{L\Omega }$ and $\hat{b}_{R\Omega }$ are operators of the
left and right-moving Rindler photons.

Here we consider an infinite chain of oscillators with the proper frequency $%
\omega _{0}=\omega e^{-a\bar{z}/c^{2}}$ in 1+1 dimension, where $\omega $ is
constant. We assume that oscillators do not move in the Rindler space and
are uniformly distributed with respect to the $\bar{z}$ coordinate. The
Rindler coordinate $\tau $ is related to the proper time for the oscillators
as $\tau _{0}=\tau e^{a\bar{z}/c^{2}}$ so that $\omega _{0}\tau _{0}=\omega
\tau $. An oscillator located at $\bar{z}$ is uniformly accelerated in
Minkowski space $(t,z)$ along the trajectory (\ref{RS1}), (\ref{RS2}).

In Minkowski space the oscillator chain moves and covers a semi-infinite
segment $z\geq c|t|$. The linear density of the oscillators in Minkowski
space goes as $|z|/(z^{2}-c^{2}t^{2})$. In a realistic situation the chain
has a finite length. In this case the speed of oscillators in the chain is
always smaller than $c$ and the linear density in Minkowski space is finite
everywhere. If the chain length is very large then it can be approximately
modeled as an infinite, which we assume in the present discussion.

We adopt the following interaction Hamiltonian between the chain of
oscillators and the scalar field 
\begin{equation*}
\hat{V}(\tau )=\hslash g\sum_{m=-\infty }^{\infty }\left( \hat{\sigma}%
_{m}e^{-i\omega \tau }+\hat{\sigma}_{m}^{\dag }e^{i\omega \tau }\right) 
\frac{\partial \hat{\Phi}(\tau ,\bar{z}_{m})}{\partial \tau },
\end{equation*}%
where $\hat{\sigma}_{m}$ and $\hat{\sigma}_{m}^{\dag }$ are lowering and
rasing operators for the oscillator $m$ located at the point $\bar{z}_{m}$,
and $g$ is the coupling constant. Next we introduce collective oscillator
operators $\hat{\sigma}_{k}$ 
\begin{equation*}
\hat{\sigma}_{k}=\frac{1}{\mathcal{N}}\sum_{m=-\infty }^{\infty }e^{-ik\bar{z%
}_{m}}\hat{\sigma}_{m},
\end{equation*}%
where $\mathcal{N}$ is a normalization factor. Then%
\begin{equation*}
\hat{\sigma}_{m}=\frac{\mathcal{N}}{2\pi \bar{\rho}}\int_{-\infty }^{\infty
}dke^{ik\bar{z}_{m}}\hat{\sigma}_{k},
\end{equation*}%
where $\bar{\rho}$ is the oscillator linear density in the Rindler space.
Plug $\hat{\sigma}_{m}$ and $\hat{\Phi}$ into the interaction Hamiltonian,
replacing the sum over $m$ with an integral over $\bar{z}$ 
\begin{equation*}
\sum_{m=-\infty }^{\infty }\rightarrow \bar{\rho}\int_{-\infty }^{\infty }d%
\bar{z},
\end{equation*}%
and integrating over $\bar{z}$ and $k$, we find%
\begin{equation*}
\hat{V}(\tau )=\frac{\hslash ga\mathcal{N}}{\sqrt{4\pi }c}\int_{0}^{\infty
}d\Omega \sqrt{\Omega }\times
\end{equation*}%
\begin{equation*}
\left( \hat{\sigma}_{-k_{\Omega }}\hat{b}_{R\Omega }e^{-i\tau \left( \frac{%
\Omega a}{c}+\omega \right) }+\hat{\sigma}_{k_{\Omega }}^{\dag }\hat{b}%
_{R\Omega }e^{-i\tau \left( \frac{\Omega a}{c}-\omega \right) }\right.
\end{equation*}%
\begin{equation*}
\left. +\hat{\sigma}_{k_{\Omega }}\hat{b}_{L\Omega }e^{-i\tau \left( \frac{%
\Omega a}{c}+\omega \right) }+\hat{\sigma}_{-k_{\Omega }}^{\dag }\hat{b}%
_{L\Omega }e^{-i\tau \left( \frac{\Omega a}{c}-\omega \right) }+h.c.\right) ,
\end{equation*}%
where $k_{\Omega }=\Omega a/c^{2}$. Finally, disregarding the
counter-rotating terms we obtain%
\begin{equation*}
\hat{V}(\tau )=\frac{\hslash ga\mathcal{N}}{\sqrt{4\pi }c}\int_{0}^{\infty
}d\Omega \sqrt{\Omega }\left( \hat{\sigma}_{k_{\Omega }}^{\dag }\hat{b}%
_{R\Omega }e^{-i\tau \left( \frac{\Omega a}{c}-\omega \right) }\right.
\end{equation*}%
\begin{equation}
\left. +\hat{\sigma}_{-k_{\Omega }}^{\dag }\hat{b}_{L\Omega }e^{-i\tau
\left( \frac{\Omega a}{c}-\omega \right) }+h.c.\right) .  \label{IH1}
\end{equation}

Equation (\ref{IH1}) shows that collective mode of the chain $\hat{\sigma}%
_{k_{\Omega }}$ is coupled with only one right-moving Rindler mode $\hat{b}%
_{R\Omega }$ and is not coupled with the left-moving Rindler photons; while
collective mode $\hat{\sigma}_{-k_{\Omega }}$ is coupled with only one
left-moving Rindler mode $\hat{b}_{L\Omega }$ and is not coupled with the
right-moving Rindler photons. Thus, in terms of the collective chain modes $%
\hat{\sigma}_{k}$ the evolution of the system is factorized. Namely, the
coupled modes $\hat{\sigma}_{k_{\Omega }}$ and $\hat{b}_{R\Omega }$ evolve
independently from the rest of the system.

\section{Classical field interacting with classical oscillator chain}

\label{KK1}

One can gain insight on the system evolution and correlations by treating
the field and oscillators classically. In the continuous limit one can model
the oscillator chain as a complex scalar field $\psi $ with a fixed natural
frequency $\omega $. If we disregard polarization the electromagnetic field
is described by a complex scalar field $\phi $. In Rindler space the action
for the field $\phi $ coupled with $\psi $ reads%
\begin{equation*}
S=\frac{1}{2}\int d\tau d\bar{z}\left[ \left\vert \frac{\partial \phi }{%
\partial \tau }\right\vert ^{2}-c^{2}\left\vert \frac{\partial \phi }{%
\partial \bar{z}}\right\vert ^{2}+\right.
\end{equation*}%
\begin{equation}
\left. \left\vert \frac{\partial \psi }{\partial \tau }\right\vert
^{2}-\omega ^{2}|\psi |^{2}+\epsilon \left( \phi \frac{\partial \psi ^{\ast }%
}{\partial \tau }+\phi ^{\ast }\frac{\partial \psi }{\partial \tau }\right) %
\right] ,
\end{equation}%
where $\epsilon $ is the coupling constant between $\phi $ and the
oscillator chain. The chain is assumed to be infinite in the Rindler space.

Variation of the action yields coupled linear equations for $\phi $ and $%
\psi $%
\begin{equation*}
\frac{\partial ^{2}\phi }{\partial \tau ^{2}}-c^{2}\frac{\partial ^{2}\phi }{%
\partial \bar{z}^{2}}-\epsilon \frac{\partial \psi }{\partial \tau }=0,
\end{equation*}%
\begin{equation*}
\frac{\partial ^{2}\psi }{\partial \tau ^{2}}+\omega ^{2}\psi +\epsilon 
\frac{\partial \phi }{\partial \tau }=0.
\end{equation*}

Since equations are linear the solution is a linear superposition of
responses to each plane-wave initially present in the field $\phi $.
Solution satisfying the initial condition 
\begin{equation*}
\phi (0,\bar{z})=e^{ik\bar{z}},\quad \psi (0,\bar{z})=0,
\end{equation*}%
and describing a collective right-moving wave reads%
\begin{equation*}
\phi (\tau ,\bar{z})=e^{-i\nu _{+}\tau +ik\bar{z}}\left( 1-\right.
\end{equation*}%
\begin{equation*}
\left. \frac{\nu _{+}\left( \omega ^{2}-\nu _{-}^{2}\right) }{\nu _{+}\left(
\omega ^{2}-\nu _{-}^{2}\right) -\nu _{-}\left( \omega ^{2}-\nu
_{+}^{2}\right) }\left[ 1-e^{i(\nu _{+}-\nu _{-})\tau }\right] \right) ,
\end{equation*}%
\begin{equation}
\psi (\tau ,\bar{z})=-\frac{i\epsilon c\omega ke^{-i\nu _{+}\tau +ik\bar{z}}%
}{\nu _{+}\left( \omega ^{2}-\nu _{-}^{2}\right) -\nu _{-}\left( \omega
^{2}-\nu _{+}^{2}\right) }\left[ 1-e^{i(\nu _{+}-\nu _{-})\tau }\right] ,
\label{cm0}
\end{equation}%
where%
\begin{equation}
\nu _{\pm }^{2}=\frac{1}{2}\left( \omega ^{2}+c^{2}k^{2}+\epsilon ^{2}\pm 
\sqrt{\left( \omega ^{2}+c^{2}k^{2}+\epsilon ^{2}\right) ^{2}-4c^{2}\omega
^{2}k^{2}}\right) .  \label{cm1}
\end{equation}%
That is field amplitudes undergo oscillations with the Rabi frequency%
\begin{equation}
\Omega =|\nu _{+}|-|\nu _{-}|.  \label{cm2}
\end{equation}%
On resonance, when $kc=\omega $ and $\epsilon \ll \omega $ we obtain%
\begin{equation*}
\nu _{\pm }\approx \omega \pm \frac{\epsilon }{2},\quad \Omega \approx
\epsilon ,
\end{equation*}%
\begin{equation*}
\psi (\tau ,\bar{z})=-\frac{i}{2}e^{-i\omega \tau +ik\bar{z}}\left(
1-e^{i\epsilon \tau }\right) ,\quad |\psi (\tau ,\bar{z})|^{2}=\sin
^{2}\left( \frac{\epsilon \tau }{2}\right) ,
\end{equation*}%
\begin{equation*}
\phi (\tau ,\bar{z})=\frac{1}{2}e^{-i\omega \tau +ik\bar{z}}\left(
1+e^{i\epsilon \tau }\right) ,\quad |\phi (\tau ,\bar{z})|^{2}=\cos
^{2}\left( \frac{\epsilon \tau }{2}\right) .
\end{equation*}%
Under these conditions, all energy initially stored in the field mode is
transferred back and forth to the oscillator chain with the Rabi frequency
given by the coupling constant $\epsilon $.

For large detuning from the resonance, namely for $|\omega -kc|\gg \epsilon $
and $\epsilon \ll \omega $ we find 
\begin{equation*}
\nu _{+}\approx \omega +\frac{\omega \epsilon ^{2}}{2\left( \omega
^{2}-c^{2}k^{2}\right) },
\end{equation*}%
\begin{equation*}
\nu _{-}\approx ck-\frac{ck\epsilon ^{2}}{2\left( \omega
^{2}-c^{2}k^{2}\right) },
\end{equation*}%
and 
\begin{equation*}
\psi (\tau ,\bar{z})=-\frac{i\epsilon ck}{\omega ^{2}-c^{2}k^{2}}e^{-i\omega
\tau +ik\bar{z}}\left[ 1-e^{i(\omega -ck)\tau }\right] ,
\end{equation*}%
\begin{equation*}
\phi (\tau ,\bar{z})=e^{-ick\tau +ik\bar{z}}\left( 1-\frac{ck\omega \epsilon
^{2}}{\left( \omega ^{2}-c^{2}k^{2}\right) ^{2}}\left[ 1-e^{-i(\omega
-ck)\tau }\right] \right) .
\end{equation*}%
In this limit, the amplitude of Rabi oscillations is reduced by a factor $%
\epsilon ^{2}c^{2}k^{2}/\left( \omega ^{2}-c^{2}k^{2}\right) ^{2}$ 
\begin{equation*}
|\psi (\tau ,\bar{z})|^{2}=\frac{4\epsilon ^{2}c^{2}k^{2}}{\left( \omega
^{2}-c^{2}k^{2}\right) ^{2}}\sin ^{2}\left( \frac{(\omega -ck)\tau }{2}%
\right) ,
\end{equation*}%
and the Rabi frequency is given by the detuning $\Omega \approx |\omega -ck|$%
.

\begin{figure}[h]
\begin{center}
\epsfig{figure=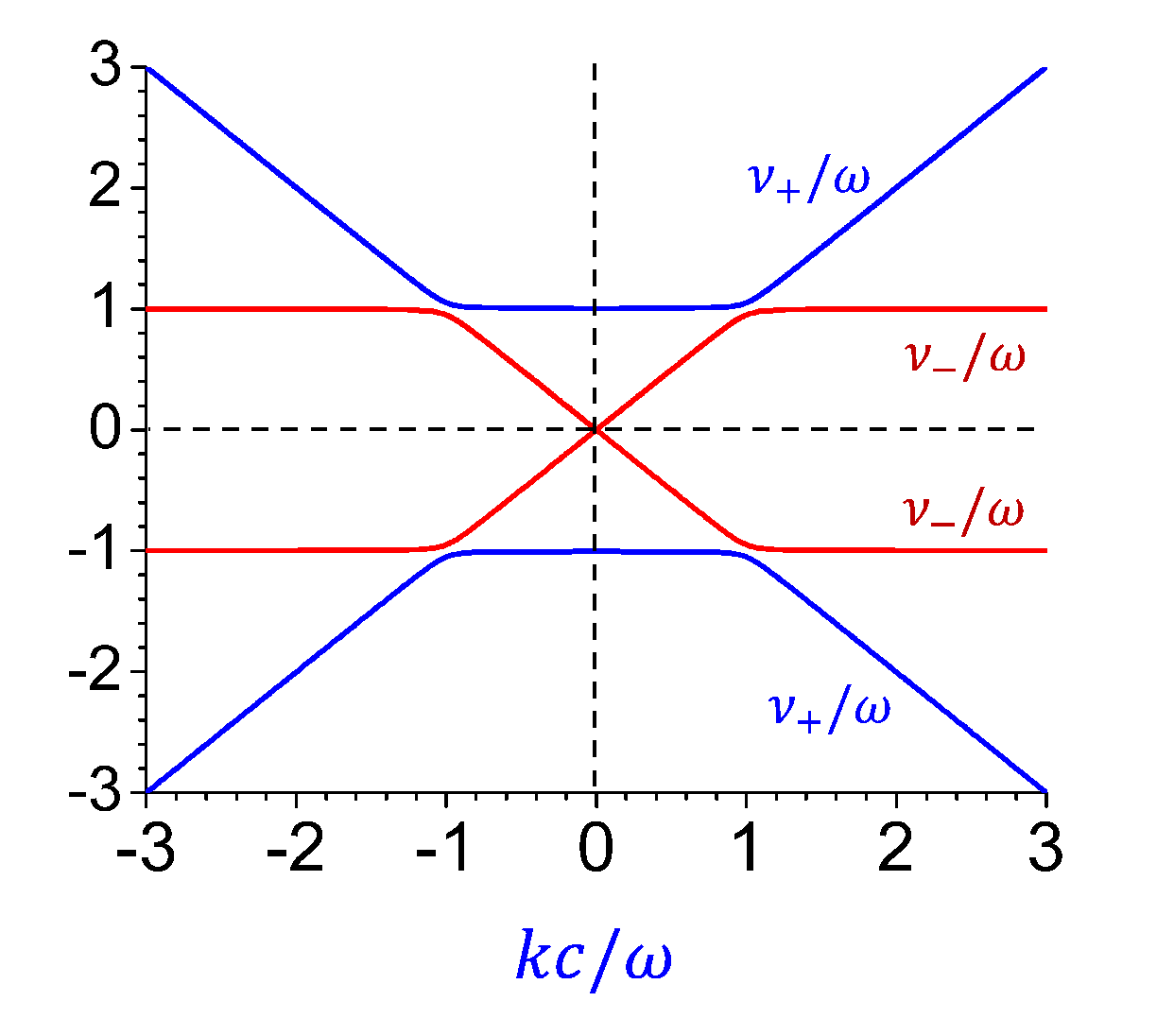, angle=0, width=9cm}
\end{center}
\caption{Collective mode frequencies $\protect\nu _{\pm }$ given by Eq. (%
\protect\ref{cm1}) as a function of $kc/\protect\omega $ for $\protect%
\epsilon =0.1\protect\omega $.}
\label{FigB1}
\end{figure}

\begin{figure}[h]
\begin{center}
\epsfig{figure=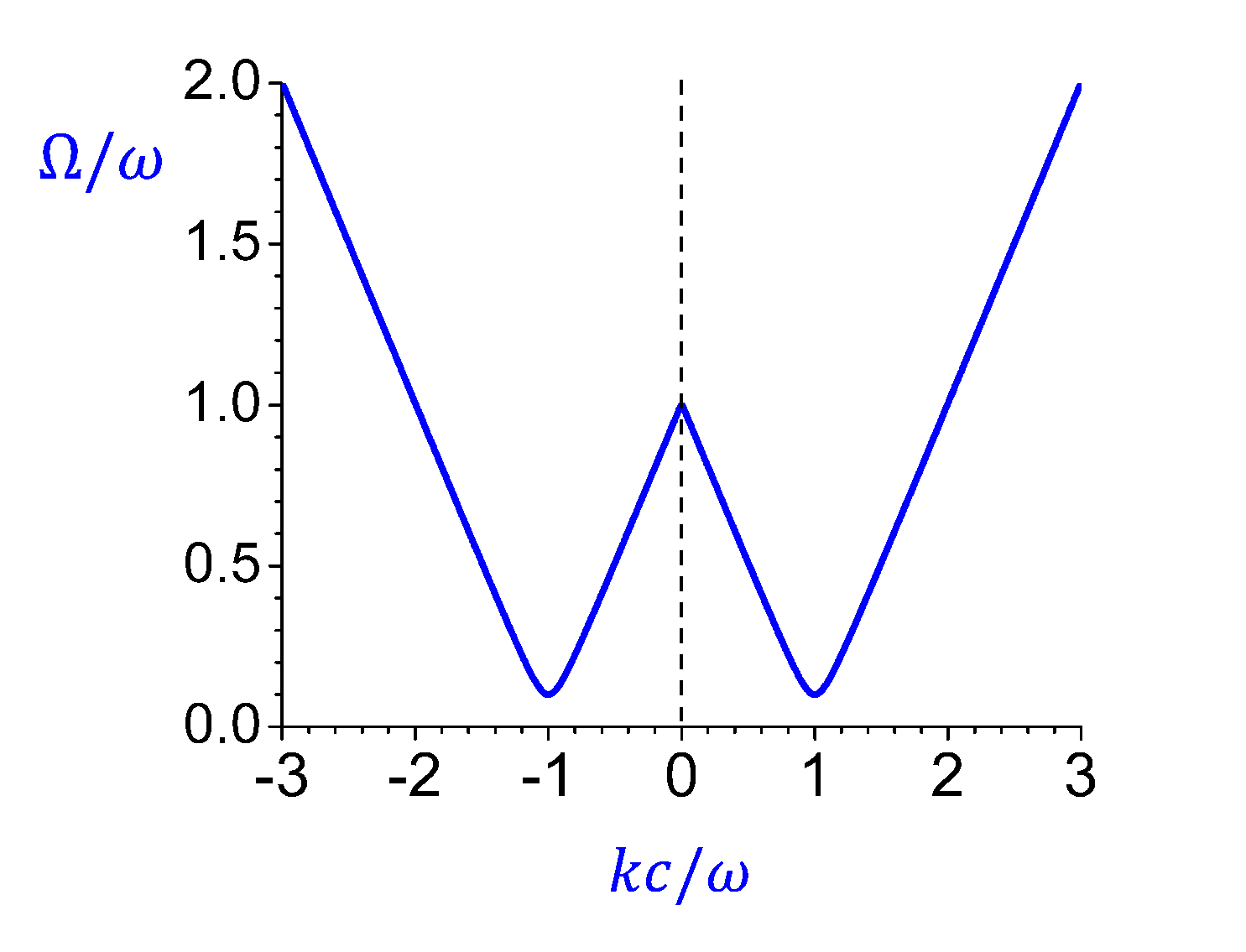, angle=0, width=9cm}
\end{center}
\caption{Rabi frequency $\Omega $ given by Eq. (\protect\ref{cm2}) as a
function of $kc/\protect\omega $ for $\protect\epsilon =0.1\protect\omega $.}
\label{FigB2}
\end{figure}

\begin{figure}[h]
\begin{center}
\epsfig{figure=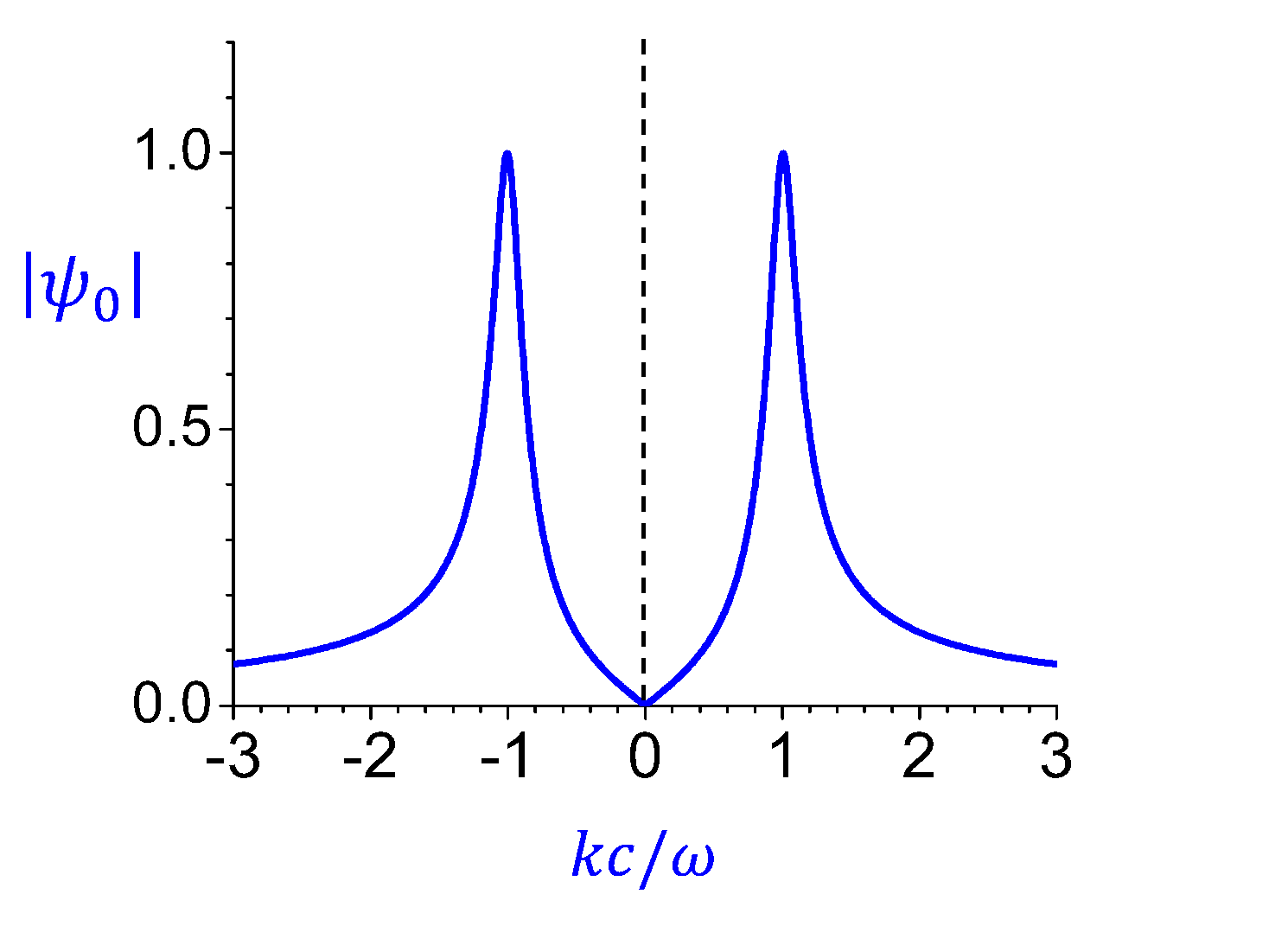, angle=0, width=9cm}
\end{center}
\caption{Amplitude of Rabi oscillations of the oscillator chain obtained
from Eq. (\protect\ref{cm0}) as a function of $kc/\protect\omega $ for $%
\protect\epsilon =0.1\protect\omega $. }
\label{FigB3}
\end{figure}

In Figs. \ref{FigB1}, \ref{FigB2} and \ref{FigB3} we plot collective mode
frequencies $\nu _{\pm }$, the Rabi frequency $\Omega $ and the amplitude of
Rabi oscillations of the oscillator chain as a function of $kc/\omega $ for $%
\epsilon =0.1\omega $. Figure \ref{FigB3} shows that the amplitude of Rabi
oscillations peaks near the resonance at which the Rabi frequency has the
smallest value (see Fig. \ref{FigB2}).

\section{State vector evolution}

\label{AA1}

Here we simplify Eq. (\ref{q0bb}) using properties of the initial state.
Taking derivative of Eq. (\ref{q0bb}) with respect to $\tau $, we have%
\begin{equation}
i\hslash \frac{\partial }{\partial \tau }\left\vert \psi (\tau
)\right\rangle =ge^{-ig\tau \left( \hat{\sigma}^{\dag }\hat{b}_{1}+\hat{%
\sigma}\hat{b}_{1}^{\dag }\right) }\hat{\sigma}^{\dag }\hat{b}_{1}\left\vert
0_{M}\right\rangle \left\vert G\right\rangle .  \label{m0b}
\end{equation}%
Next we prove an identity%
\begin{equation}
\hat{b}_{1}e^{\gamma \hat{b}_{1}^{\dagger }\hat{b}_{2}^{\dagger }}=e^{\gamma 
\hat{b}_{1}^{\dagger }\hat{b}_{2}^{\dagger }}\hat{b}_{1}+\gamma \hat{b}%
_{2}^{\dagger }e^{\gamma \hat{b}_{1}^{\dagger }\hat{b}_{2}^{\dagger }}.
\label{q1}
\end{equation}%
Introducing operator 
\begin{equation*}
\hat{Q}(\gamma )=\hat{b}_{1}e^{\gamma \hat{b}_{1}^{\dagger }\hat{b}%
_{2}^{\dagger }}-e^{\gamma \hat{b}_{1}^{\dagger }\hat{b}_{2}^{\dagger }}\hat{%
b}_{1},
\end{equation*}%
and taking derivative over $\gamma $, we obtain a differential equation for $%
\hat{Q}(\gamma )$%
\begin{equation}
\frac{\partial \hat{Q}}{\partial \gamma }=\hat{b}_{2}^{\dagger }e^{\gamma 
\hat{b}_{1}^{\dagger }\hat{b}_{2}^{\dagger }}+\hat{b}_{1}^{\dagger }\hat{b}%
_{2}^{\dagger }\hat{Q}.  \label{m2}
\end{equation}%
Solution of this equation subject to the condition $\hat{Q}(0)=0$ is%
\begin{equation*}
\hat{Q}(\gamma )=\gamma \hat{b}_{2}^{\dagger }e^{\gamma \hat{b}_{1}^{\dagger
}\hat{b}_{2}^{\dagger }},
\end{equation*}%
which coincides with Eq. (\ref{q1}). Using identity (\ref{q1}) and the
formula for the Minkowski vacuum $\left\vert 0_{M}\right\rangle $ in terms
of the Rindler vacuum $\left\vert 0_{R}\right\rangle $%
\begin{equation}
\left\vert 0_{M}\right\rangle =\sqrt{1-\gamma ^{2}}e^{\gamma \hat{b}%
_{1}^{\dagger }\hat{b}_{2}^{\dagger }}\left\vert 0_{R}\right\rangle ,
\label{xx2}
\end{equation}
one can rewrite Eq. (\ref{m0b}) as%
\begin{equation}
i\hslash \frac{\partial }{\partial \tau }\left\vert \psi (\tau
)\right\rangle =g\gamma \hat{b}_{2}^{\dagger }e^{-ig\tau \left( \hat{\sigma}%
^{\dag }\hat{b}_{1}+\hat{\sigma}\hat{b}_{1}^{\dag }\right) }\hat{\sigma}%
^{\dag }\left\vert 0_{M}\right\rangle \left\vert G\right\rangle .
\label{m3b}
\end{equation}

Next we calculate the commutator 
\begin{equation*}
\hat{W}=e^{s\hat{\sigma}^{\dag }\hat{b}_{1}-s^{\ast }\hat{\sigma}\hat{b}%
_{1}^{\dag }}\hat{\sigma}^{\dag }-\hat{\sigma}^{\dag }e^{s\hat{\sigma}^{\dag
}\hat{b}_{1}-s^{\ast }\hat{\sigma}\hat{b}_{1}^{\dag }},
\end{equation*}%
where $s$ is a complex number. For the present problem $s=-ig\tau $.
Introducing operator 
\begin{equation*}
\hat{W}(t)=e^{\left( s\hat{\sigma}^{\dag }\hat{b}_{1}-s^{\ast }\hat{\sigma}%
\hat{b}_{1}^{\dag }\right) t}\hat{\sigma}^{\dag }-\hat{\sigma}^{\dag
}e^{\left( s\hat{\sigma}^{\dag }\hat{b}_{1}-s^{\ast }\hat{\sigma}\hat{b}%
_{1}^{\dag }\right) t},
\end{equation*}%
and taking derivative over $t$, we obtain differential equation for $\hat{W}%
(t)$%
\begin{equation*}
\frac{\partial \hat{W}(t)}{\partial t}=\left( s\hat{\sigma}^{\dag }\hat{b}%
_{1}-s^{\ast }\hat{\sigma}\hat{b}_{1}^{\dag }\right) \hat{W}(t)-s^{\ast }%
\hat{b}_{1}^{\dag }e^{\left( s\hat{\sigma}^{\dag }\hat{b}_{1}-s^{\ast }\hat{%
\sigma}\hat{b}_{1}^{\dag }\right) t}.
\end{equation*}%
We look for solution of this equation in the form%
\begin{equation*}
\hat{W}(t)=\hat{P}(t)e^{\left( s\hat{\sigma}^{\dag }\hat{b}_{1}-s^{\ast }%
\hat{\sigma}\hat{b}_{1}^{\dag }\right) t},
\end{equation*}%
which yields the following equation for the operator $\hat{P}(t)$%
\begin{equation*}
\frac{\partial \hat{P}(t)}{\partial t}=-s^{\ast }\hat{b}_{1}^{\dag }+
\end{equation*}%
\begin{equation*}
\left( s\hat{\sigma}^{\dag }\hat{b}_{1}-s^{\ast }\hat{\sigma}\hat{b}%
_{1}^{\dag }\right) \hat{P}(t)-\hat{P}(t)\left( s\hat{\sigma}^{\dag }\hat{b}%
_{1}-s^{\ast }\hat{\sigma}\hat{b}_{1}^{\dag }\right) .
\end{equation*}%
Solution of this differential equation, satisfying condition $\hat{P}(0)=0$,
is%
\begin{equation*}
\hat{P}(t)=\left( \cos (|s|t)-1\right) \hat{\sigma}^{\dag }-\frac{s^{\ast }}{%
|s|}\sin (|s|t)\hat{b}_{1}^{\dag },
\end{equation*}%
and, therefore, 
\begin{equation}
\hat{W}(t)=\left( \left( \cos (|s|t)-1\right) \hat{\sigma}^{\dag }-\frac{%
s^{\ast }}{|s|}\sin (|s|t)\hat{b}_{1}^{\dag }\right) e^{\left( s\hat{\sigma}%
^{\dag }\hat{b}_{1}-s^{\ast }\hat{\sigma}\hat{b}_{1}^{\dag }\right) t}.
\label{m4}
\end{equation}%
Using Eq. (\ref{m4}), we obtain%
\begin{equation}
e^{s\hat{\sigma}^{\dag }\hat{b}_{1}-s^{\ast }\hat{\sigma}\hat{b}_{1}^{\dag }}%
\hat{\sigma}^{\dag }=\left( \cos (|s|)\hat{\sigma}^{\dag }-\frac{s^{\ast }}{%
|s|}\sin (|s|)\hat{b}_{1}^{\dag }\right) e^{s\hat{\sigma}^{\dag }\hat{b}%
_{1}-s^{\ast }\hat{\sigma}\hat{b}_{1}^{\dag }}.  \label{m5bb}
\end{equation}%
Plug Eq. (\ref{m5bb}) in Eq. (\ref{m3b}), yields 
\begin{equation}
i\hslash \frac{\partial }{\partial \tau }\left\vert \psi (\tau
)\right\rangle =-g\gamma \hat{b}_{2}^{\dagger }\left( i\sin (g\tau )\hat{b}%
_{1}^{\dag }-\cos (g\tau )\hat{\sigma}^{\dag }\right) \left\vert \psi (\tau
)\right\rangle .  \label{m6b}
\end{equation}%
Solution of the differential Eq. (\ref{m6b}) for the field state vector is
given by%
\begin{equation*}
\left\vert \psi (\tau )\right\rangle =e^{\gamma \hat{b}_{2}^{\dagger }\left(
\left( \cos (g\tau )-1\right) \hat{b}_{1}^{\dag }-i\sin (g\tau )\hat{\sigma}%
^{\dag }\right) }\left\vert G\right\rangle \left\vert 0_{M}\right\rangle .
\end{equation*}

\section{Reduced density operator for Rindler photons}

\label{AA2}

Here we calculate the reduced density operator for the Rindler photons in
the mode $\hat{b}_{1}$. Eq. (\ref{m7c}) for the state vector can be written
as%
\begin{equation*}
\left\vert \psi (\tau )\right\rangle =\sqrt{1-\gamma ^{2}}\times
\end{equation*}%
\begin{equation*}
\sum_{n=0}^{\infty }\frac{\gamma ^{n}}{\sqrt{n!}}\left( \cos (g\tau )\hat{b}%
_{1}^{\dag }-i\sin (g\tau )\hat{\sigma}^{\dag }\right) ^{n}\left\vert
0n\right\rangle \left\vert G\right\rangle ,
\end{equation*}%
where $\left\vert 0n\right\rangle $ is a state with $0$ photons in the mode $%
\hat{b}_{1}$ and $n$ photons in the mode $\hat{b}_{2}$. Tracing over the
mode $\hat{b}_{2}$, we obtain for the reduced density operator%
\begin{equation*}
\hat{\rho}_{b_{1},\sigma }=\left( 1-\gamma ^{2}\right) \sum_{n=0}^{\infty }%
\frac{\gamma ^{2n}}{n!}\left( \cos (g\tau )\hat{b}_{1}^{\dag }-i\sin (g\tau )%
\hat{\sigma}^{\dag }\right) ^{n}\left\vert G\right\rangle \left\vert
0\right\rangle
\end{equation*}%
\begin{equation}
\left\langle 0\right\vert \left\langle G\right\vert \left( \cos (g\tau )\hat{%
b}_{1}+i\sin (g\tau )\hat{\sigma}\right) ^{n},  \label{P1}
\end{equation}%
where $\left\vert 0\right\rangle $ is a state with $0$ Rindler photons $\hat{%
b}_{1}$. Using formula for the binomial expansion%
\begin{equation*}
\left( x+y\right) ^{n}=\sum_{k=0}^{n}\frac{n!}{k!(n-k)!}x^{n-k}y^{k},
\end{equation*}%
one can write this equation as%
\begin{equation*}
\hat{\rho}_{b_{1},\sigma }=\left( 1-\gamma ^{2}\right) \sum_{n=0}^{\infty
}\sum_{k=0}^{n}\frac{\left( -i\right) ^{k}\gamma ^{2n}\sin ^{k}(g\tau )\cos
^{n-k}(g\tau )}{k!(n-k)!}
\end{equation*}%
\begin{equation*}
\hat{b}_{1}^{\dag n-k}\hat{\sigma}^{\dag k}\left\vert G\right\rangle
\left\vert 0\right\rangle \left\langle 0\right\vert \left\langle
G\right\vert \sum_{m=0}^{n}\frac{i^{m}n!\sin ^{m}(g\tau )\cos ^{n-m}(g\tau )%
}{m!(n-m)!}\hat{b}_{1}^{n-m}\hat{\sigma}^{m}.
\end{equation*}%
Tracing over the oscillator, we obtain the reduced density operator for the
field in the mode $\hat{b}_{1}$%
\begin{equation*}
\hat{\rho}_{b_{1}}=\left( 1-\gamma ^{2}\right) \sum_{m=0}^{\infty
}\sum_{n=m}^{\infty }\frac{\gamma ^{2n}n!\sin ^{2(n-m)}(g\tau )\cos
^{2m}(g\tau )}{(n-m)!m!}\left\vert m\right\rangle \left\langle m\right\vert
\end{equation*}%
\begin{equation}
=\left( 1-\gamma ^{2}\right) \sum_{m=0}^{\infty }\frac{\gamma ^{2m}\cos
^{2m}(g\tau )}{\left( 1-\gamma ^{2}\sin ^{2}(g\tau )\right) ^{m+1}}%
\left\vert m\right\rangle \left\langle m\right\vert ,  \label{w2}
\end{equation}%
where $\left\vert m\right\rangle $ is a state with $m$ Rindler photons $\hat{%
b}_{1}$, and we used%
\begin{equation*}
\sum_{n=m}^{\infty }\frac{\gamma ^{2n}n!}{(n-m)!}=\frac{\gamma ^{2m}m!}{%
\left( 1-\gamma ^{2}\right) ^{m+1}.}
\end{equation*}

\section{State vector in terms of Unruh-Minkowski photons}

\label{AA3}

First, we calculate $e^{-A\hat{b}_{1}^{\dagger }\hat{b}_{2}^{\dagger
}}\left\vert 0_{M}\right\rangle $, where $A$ is a complex number. For this
purpose, we introduce a state vector%
\begin{equation*}
F(t)=e^{\left( \alpha \hat{a}_{1}^{\dagger }\hat{a}_{2}^{\dagger }+\beta
\left( \hat{a}_{1}^{\dagger }\hat{a}_{1}+\hat{a}_{2}\hat{a}_{2}^{\dagger
}\right) +\lambda \hat{a}_{2}\hat{a}_{1}\right) t}e^{f(t)\hat{a}%
_{1}^{\dagger }\hat{a}_{2}^{\dagger }}\left\vert 0_{M}\right\rangle ,
\end{equation*}%
where $f(t)$ is a function of $t$ which we obtain later. In the following,
we will use identities%
\begin{equation}
\hat{a}_{1}e^{f\hat{a}_{1}^{\dagger }\hat{a}_{2}^{\dagger }}=e^{f\hat{a}%
_{1}^{\dagger }\hat{a}_{2}^{\dagger }}\hat{a}_{1}+fe^{f\hat{a}_{1}^{\dagger }%
\hat{a}_{2}^{\dagger }}\hat{a}_{2}^{\dagger },  \label{z1}
\end{equation}%
\begin{equation}
\hat{a}_{2}e^{f\hat{a}_{1}^{\dagger }\hat{a}_{2}^{\dagger }}=e^{f\hat{a}%
_{1}^{\dagger }\hat{a}_{2}^{\dagger }}\hat{a}_{2}+fe^{f\hat{a}_{1}^{\dagger }%
\hat{a}_{2}^{\dagger }}\hat{a}_{1}^{\dagger },  \label{z2}
\end{equation}%
which have been proven previously (see Eq. (\ref{q1})). Taking derivative
with respect to $t$, we have%
\begin{equation*}
\frac{d}{dt}F(t)=e^{\left( \alpha \hat{a}_{1}^{\dagger }\hat{a}_{2}^{\dagger
}+\beta \left( \hat{a}_{1}^{\dagger }\hat{a}_{1}+\hat{a}_{2}\hat{a}%
_{2}^{\dagger }\right) +\gamma \hat{a}_{2}\hat{a}_{1}\right) t}\times
\end{equation*}%
\begin{equation*}
\left[ \left( \alpha +\dot{f}\right) \hat{a}_{1}^{\dagger }\hat{a}%
_{2}^{\dagger }+\beta \left( \hat{a}_{1}^{\dagger }\hat{a}_{1}+\hat{a}_{2}%
\hat{a}_{2}^{\dagger }\right) +\gamma \hat{a}_{2}\hat{a}_{1}\right] e^{f\hat{%
a}_{1}^{\dagger }\hat{a}_{2}^{\dagger }}\left\vert 0_{M}\right\rangle .
\end{equation*}

Using identities (\ref{z1}) and (\ref{z2}), we find%
\begin{equation*}
\hat{a}_{2}\hat{a}_{1}e^{f\hat{a}_{1}^{\dagger }\hat{a}_{2}^{\dagger }}=e^{f%
\hat{a}_{1}^{\dagger }\hat{a}_{2}^{\dagger }}\hat{a}_{2}\hat{a}_{1}+fe^{f%
\hat{a}_{1}^{\dagger }\hat{a}_{2}^{\dagger }}\left( \hat{a}_{1}^{\dagger }%
\hat{a}_{1}+\hat{a}_{2}\hat{a}_{2}^{\dagger }\right)
\end{equation*}%
\begin{equation*}
+f^{2}e^{f\hat{a}_{1}^{\dagger }\hat{a}_{2}^{\dagger }}\hat{a}_{1}^{\dagger }%
\hat{a}_{2}^{\dagger },
\end{equation*}%
\begin{equation*}
\hat{a}_{1}^{\dagger }\hat{a}_{1}e^{f\hat{a}_{1}^{\dagger }\hat{a}%
_{2}^{\dagger }}=e^{f\hat{a}_{1}^{\dagger }\hat{a}_{2}^{\dagger }}\hat{a}%
_{1}^{\dagger }\hat{a}_{1}+fe^{f\hat{a}_{1}^{\dagger }\hat{a}_{2}^{\dagger }}%
\hat{a}_{1}^{\dagger }\hat{a}_{2}^{\dagger },
\end{equation*}%
\begin{equation*}
\hat{a}_{2}\hat{a}_{2}^{\dagger }e^{f\hat{a}_{1}^{\dagger }\hat{a}%
_{2}^{\dagger }}=e^{f\hat{a}_{1}^{\dagger }\hat{a}_{2}^{\dagger }}\hat{a}_{2}%
\hat{a}_{2}^{\dagger }+fe^{f\hat{a}_{1}^{\dagger }\hat{a}_{2}^{\dagger }}%
\hat{a}_{1}^{\dagger }\hat{a}_{2}^{\dagger }.
\end{equation*}%
As a result,%
\begin{equation*}
\frac{d}{dt}F(t)=e^{\left( \alpha \hat{a}_{1}^{\dagger }\hat{a}_{2}^{\dagger
}+\beta \left( \hat{a}_{1}^{\dagger }\hat{a}_{1}+\hat{a}_{2}\hat{a}%
_{2}^{\dagger }\right) +\lambda \hat{a}_{2}\hat{a}_{1}\right) t}e^{f\hat{a}%
_{1}^{\dagger }\hat{a}_{2}^{\dagger }}\times
\end{equation*}%
\begin{equation*}
\left[ \left( \alpha +\dot{f}+2\beta f+\lambda f^{2}\right) \hat{a}%
_{1}^{\dagger }\hat{a}_{2}^{\dagger }+\beta +\lambda f\right] \left\vert
0_{M}\right\rangle .
\end{equation*}%
We choose $f(t)$ as a solution of the equation 
\begin{equation*}
\alpha +\dot{f}+2\beta f+\lambda f^{2}=0,
\end{equation*}%
that is%
\begin{equation}
\beta +\lambda f(t)=-\mu \tan \left( \mu t+C\right) ,  \label{z3}
\end{equation}%
where%
\begin{equation*}
\mu =\sqrt{\alpha \lambda -\beta ^{2}}.
\end{equation*}

We fix the integration constant $C$ by requiring $f(0)=0$, which yields 
\begin{equation*}
\tan \left( C\right) =-\frac{\beta }{\mu }.
\end{equation*}%
Then $F(t)$ obeys the differential equation 
\begin{equation*}
\frac{d}{dt}F(t)=-\mu \tan \left( \mu t+C\right) F(t)\text{,}
\end{equation*}%
subject to the condition $F(0)=\left\vert 0_{M}\right\rangle $. Solution of
this equation reads%
\begin{equation*}
F(t)=\left( \cos \left( \mu t\right) +\frac{\beta }{\mu }\sin \left( \mu
t\right) \right) \left\vert 0_{M}\right\rangle .
\end{equation*}%
Taking $t=1$, we have%
\begin{equation*}
F(1)=e^{\alpha \hat{a}_{1}^{\dagger }\hat{a}_{2}^{\dagger }+\beta \left( 
\hat{a}_{1}^{\dagger }\hat{a}_{1}+\hat{a}_{2}\hat{a}_{2}^{\dagger }\right)
+\lambda \hat{a}_{2}\hat{a}_{1}}e^{f(1)\hat{a}_{1}^{\dagger }\hat{a}%
_{2}^{\dagger }}\left\vert 0_{M}\right\rangle =
\end{equation*}%
\begin{equation}
\left( \cos \left( \mu \right) +\frac{\beta }{\mu }\sin \left( \mu \right)
\right) \left\vert 0_{M}\right\rangle ,
\end{equation}%
or%
\begin{equation*}
e^{-\alpha \hat{a}_{1}^{\dagger }\hat{a}_{2}^{\dagger }-\beta \left( \hat{a}%
_{1}^{\dagger }\hat{a}_{1}+\hat{a}_{2}\hat{a}_{2}^{\dagger }\right) -\lambda 
\hat{a}_{2}\hat{a}_{1}}\left\vert 0_{M}\right\rangle =
\end{equation*}%
\begin{equation}
\frac{1}{\cos \left( \mu \right) +\frac{\beta }{\mu }\sin \left( \mu \right) 
}e^{f(1)\hat{a}_{1}^{\dagger }\hat{a}_{2}^{\dagger }}\left\vert
0_{M}\right\rangle .
\end{equation}%
For the present problem%
\begin{equation*}
\alpha =\frac{A}{1-\gamma ^{2}},\quad \beta =\frac{A\gamma }{1-\gamma ^{2}}%
,\quad \lambda =\frac{A\gamma ^{2}}{1-\gamma ^{2}}.
\end{equation*}%
Thus,%
\begin{equation*}
\alpha \lambda -\beta ^{2}=0,
\end{equation*}%
and Eq. (\ref{z3}) gives%
\begin{equation*}
f(1)=-\frac{\alpha }{1+\beta }=-\frac{A}{1+A\gamma -\gamma ^{2}}.
\end{equation*}%
Finally, we obtain%
\begin{equation}
e^{-A\hat{b}_{1}^{\dagger }\hat{b}_{2}^{\dagger }}\left\vert
0_{M}\right\rangle =\frac{1-\gamma ^{2}}{1+A\gamma -\gamma ^{2}}e^{-\frac{A}{%
1+A\gamma -\gamma ^{2}}\hat{a}_{1}^{\dagger }\hat{a}_{2}^{\dagger
}}\left\vert 0_{M}\right\rangle .  \label{z4}
\end{equation}

Eqs. (\ref{xx1}) and (\ref{z4}) yield that the state vector of the system in
terms of the Unruh-Minkowski photons is given by%
\begin{equation*}
\left\vert \psi (\tau )\right\rangle =e^{-i\gamma \sin (g\tau )\hat{b}%
_{2}^{\dagger }\hat{\sigma}^{\dag }}e^{\gamma \left( \cos (g\tau )-1\right) 
\hat{b}_{2}^{\dagger }\hat{b}_{1}^{\dag }}\left\vert G\right\rangle
\left\vert 0_{M}\right\rangle =
\end{equation*}%
\begin{equation*}
=\frac{1-\gamma ^{2}}{1-\gamma ^{2}\cos (g\tau )}\times
\end{equation*}%
\begin{equation}
e^{-i\frac{\gamma }{\sqrt{1-\gamma ^{2}}}\sin (g\tau )\left( \hat{a}%
_{2}^{\dagger }+\gamma \hat{a}_{1}\right) \hat{\sigma}^{\dag }}e^{-\frac{%
\gamma \left( 1-\cos (g\tau )\right) }{1-\gamma ^{2}\cos (g\tau )}\hat{a}%
_{1}^{\dagger }\hat{a}_{2}^{\dagger }}\left\vert G\right\rangle \left\vert
0_{M}\right\rangle .  \label{x3}
\end{equation}%
Next we prove an identity%
\begin{equation}
e^{\alpha \hat{a}_{1}\hat{\sigma}^{\dag }}e^{\beta \hat{a}_{1}^{\dagger }%
\hat{a}_{2}^{\dagger }}=e^{\alpha \beta \hat{a}_{2}^{\dagger }\hat{\sigma}%
^{\dag }}e^{\beta \hat{a}_{1}^{\dagger }\hat{a}_{2}^{\dagger }}e^{\alpha 
\hat{a}_{1}\hat{\sigma}^{\dag }}.  \label{x4}
\end{equation}%
Introducing operator 
\begin{equation*}
\hat{P}(\alpha )=e^{\alpha \hat{a}_{1}\hat{\sigma}^{\dag }}e^{\beta \hat{a}%
_{1}^{\dagger }\hat{a}_{2}^{\dagger }}e^{-\alpha \hat{a}_{1}\hat{\sigma}%
^{\dag }},
\end{equation*}%
and taking derivative with respect to $\alpha $, we obtain 
\begin{equation*}
\frac{d}{d\alpha }\hat{P}(\alpha )=\hat{\sigma}^{\dag }e^{\alpha \hat{a}_{1}%
\hat{\sigma}^{\dag }}\left( \hat{a}_{1}e^{\beta \hat{a}_{1}^{\dagger }\hat{a}%
_{2}^{\dagger }}-e^{\beta \hat{a}_{1}^{\dagger }\hat{a}_{2}^{\dagger }}\hat{a%
}_{1}\right) e^{-\alpha \hat{a}_{1}\hat{\sigma}^{\dag }}.
\end{equation*}%
Using Eq. (\ref{z1}), we have%
\begin{equation*}
\frac{d}{d\alpha }\hat{P}(\alpha )=\beta \hat{a}_{2}^{\dagger }\hat{\sigma}%
^{\dag }\hat{P}(\alpha ).
\end{equation*}%
Solution of this differential equation, subject to the condition $\hat{P}%
(0)=e^{\beta \hat{a}_{1}^{\dagger }\hat{a}_{2}^{\dagger }}$, is 
\begin{equation*}
\hat{P}(\alpha )=e^{\alpha \beta \hat{a}_{2}^{\dagger }\hat{\sigma}^{\dag
}}e^{\beta \hat{a}_{1}^{\dagger }\hat{a}_{2}^{\dagger }},
\end{equation*}%
which proves the identity (\ref{x4}).

Using Eq. (\ref{x4}), one can write the state vector of the system as%
\begin{equation*}
\left\vert \psi (\tau )\right\rangle =\frac{1-\gamma ^{2}}{1-\gamma ^{2}\cos
(g\tau )}e^{-i\gamma \sqrt{1-\gamma ^{2}}\frac{\sin (g\tau )}{1-\gamma
^{2}\cos (g\tau )}\hat{a}_{2}^{\dagger }\hat{\sigma}^{\dag }}
\end{equation*}%
\begin{equation*}
\times e^{-\frac{\gamma \left( 1-\cos (g\tau )\right) }{1-\gamma ^{2}\cos
(g\tau )}\hat{a}_{1}^{\dagger }\hat{a}_{2}^{\dagger }}\left\vert
G\right\rangle \left\vert 0_{M}\right\rangle .
\end{equation*}

\section{Simplification of Eq. (\protect\ref{q0})}

\label{AA4}

Here we simplify Eq. (\ref{q0}) using properties of the initial state.
Taking derivative of Eq. (\ref{q0}) with respect to $\tau $, we obtain%
\begin{equation*}
i\hslash \frac{\partial }{\partial \tau }\left\vert \psi (\tau
)\right\rangle =
\end{equation*}%
\begin{equation}
ge^{-ig\tau \left( \hat{\sigma}_{1}^{\dag }\hat{b}_{1}+\hat{\sigma}_{1}\hat{b%
}_{1}^{\dag }+\hat{\sigma}_{2}^{\dag }\hat{b}_{2}+\hat{\sigma}_{2}\hat{b}%
_{2}^{\dag }\right) }\left( \hat{\sigma}_{1}^{\dag }\hat{b}_{1}+\hat{\sigma}%
_{2}^{\dag }\hat{b}_{2}\right) \left\vert 0_{M}\right\rangle \left\vert
G\right\rangle .  \label{m0}
\end{equation}%
Using identity (\ref{q1}) and the formula for the Minkowski vacuum (\ref{xx2}%
), one can rewrite Eq. (\ref{m0}) as%
\begin{equation*}
i\hslash \frac{\partial }{\partial \tau }\left\vert \psi (\tau
)\right\rangle =
\end{equation*}%
\begin{equation}
g\gamma e^{-ig\tau \left( \hat{\sigma}_{1}^{\dag }\hat{b}_{1}+\hat{\sigma}%
_{1}\hat{b}_{1}^{\dag }+\hat{\sigma}_{2}^{\dag }\hat{b}_{2}+\hat{\sigma}_{2}%
\hat{b}_{2}^{\dag }\right) }\left( \hat{\sigma}_{1}^{\dag }\hat{b}%
_{2}^{\dagger }+\hat{\sigma}_{2}^{\dag }\hat{b}_{1}^{\dagger }\right)
\left\vert 0_{M}\right\rangle \left\vert G\right\rangle .  \label{m3}
\end{equation}

Next we use identities%
\begin{equation*}
e^{s\hat{\sigma}_{1}^{\dag }\hat{b}_{1}-s^{\ast }\hat{\sigma}_{1}\hat{b}%
_{1}^{\dag }}\hat{\sigma}_{1}^{\dag }=\left( \cos (|s|)\hat{\sigma}%
_{1}^{\dag }-\frac{s^{\ast }}{|s|}\sin (|s|)\hat{b}_{1}^{\dag }\right) e^{s%
\hat{\sigma}_{1}^{\dag }\hat{b}_{1}-s^{\ast }\hat{\sigma}_{1}\hat{b}%
_{1}^{\dag }},
\end{equation*}%
\begin{equation*}
e^{s\hat{\sigma}_{2}^{\dag }\hat{b}_{2}-s^{\ast }\hat{\sigma}_{2}\hat{b}%
_{2}^{\dag }}\hat{b}_{2}^{\dagger }=\left( \cos (|s|)\hat{b}_{2}^{\dagger }+%
\frac{s}{|s|}\sin (|s|)\hat{\sigma}_{2}^{\dag }\right) e^{s\hat{\sigma}%
_{2}^{\dag }\hat{b}_{2}-s^{\ast }\hat{\sigma}_{2}\hat{b}_{2}^{\dag }},
\end{equation*}%
\begin{equation*}
e^{s\hat{\sigma}_{2}^{\dag }\hat{b}_{2}-s^{\ast }\hat{\sigma}_{2}\hat{b}%
_{2}^{\dag }}\hat{\sigma}_{2}^{\dag }=\left( \cos (|s|)\hat{\sigma}%
_{2}^{\dag }-\frac{s^{\ast }}{|s|}\sin (|s|)\hat{b}_{2}^{\dag }\right) e^{s%
\hat{\sigma}_{2}^{\dag }\hat{b}_{2}-s^{\ast }\hat{\sigma}_{2}\hat{b}%
_{2}^{\dag }},
\end{equation*}%
\begin{equation*}
e^{s\hat{\sigma}_{1}^{\dag }\hat{b}_{1}-s^{\ast }\hat{\sigma}_{1}\hat{b}%
_{1}^{\dag }}\hat{b}_{1}^{\dagger }=\left( \cos (|s|)\hat{b}_{1}^{\dagger }+%
\frac{s}{|s|}\sin (|s|)\hat{\sigma}_{1}^{\dag }\right) e^{s\hat{\sigma}%
_{1}^{\dag }\hat{b}_{1}-s^{\ast }\hat{\sigma}_{1}\hat{b}_{1}^{\dag }},
\end{equation*}%
where $s$ is a complex number. For the present problem $s=-ig\tau $. Plug
these equations in Eq. (\ref{m3}), yields%
\begin{equation*}
\frac{\partial }{\partial \tau }\left\vert \psi (\tau )\right\rangle
=-g\gamma \left( \sin (2g\tau )\left( \hat{b}_{1}^{\dag }\hat{b}%
_{2}^{\dagger }+\hat{\sigma}_{1}^{\dag }\hat{\sigma}_{2}^{\dag }\right)
\right.
\end{equation*}%
\begin{equation}
+\left. i\cos (2g\tau )\left( \hat{b}_{2}^{\dag }\hat{\sigma}_{1}^{\dag }+%
\hat{b}_{1}^{\dag }\hat{\sigma}_{2}^{\dag }\right) \right) \left\vert \psi
(\tau )\right\rangle .  \label{m6}
\end{equation}

Solution of the differential Eq. (\ref{m6}) for the system's state vector is%
\begin{equation*}
\left\vert \psi (\tau )\right\rangle =\exp \left[ \frac{\gamma }{2}\left(
\left( \cos (2g\tau )-1\right) \left( \hat{b}_{1}^{\dag }\hat{b}%
_{2}^{\dagger }+\hat{\sigma}_{1}^{\dag }\hat{\sigma}_{2}^{\dag }\right)
\right. \right.
\end{equation*}%
\begin{equation*}
-\left. \left. i\sin (2g\tau )\left( \hat{b}_{2}^{\dag }\hat{\sigma}%
_{1}^{\dag }+\hat{b}_{1}^{\dag }\hat{\sigma}_{2}^{\dag }\right) \right) %
\right] \left\vert G\right\rangle \left\vert 0_{M}\right\rangle .
\end{equation*}

\end{document}